\documentclass[prd,tightenlines,twoside,secnumarabic,
               onecolumn,floatfix,nofootinbib,showpacs,11pt]{revtex4}

\usepackage[english]{babel}
\usepackage{amsmath,amssymb,bm,slashed}
\usepackage{graphicx}
\usepackage[sort&compress]{natbib}
\usepackage{xcolor}
\usepackage[normalem]{ulem}
\usepackage{hyperref}
\usepackage{cleveref}
\usepackage{subfigure} 
\definecolor{red}{rgb}{1.0, 0, 0}

\allowdisplaybreaks

\bibpunct{[}{]}{,}{n}{}{,}

\newcommand{\ev}[1]{\ensuremath{\left\langle #1 %
                     \right\rangle}} 
\newcommand{\tr}{\text{tr}}

\newcommand{\parenbar}[1]{\overset{
            \raisebox{-0.15em}{\scalebox{.4}{\textbf{(}}}
            \raisebox{-0.3em}{{\hspace{.03em}--\hspace{.05em}}}
            \raisebox{-0.15em}{\scalebox{.4}{\textbf{)}}}} {#1}}

\renewcommand{\vec}[1]{{\mathbf{#1}}}

\newcommand{\beq}{\begin{equation}}
\newcommand{\eeq}{\end{equation}}

\newcommand{\MSbar}[0]{\overline{\text{MS}}}

\begin{document}

\title{From gamma ray line signals of dark matter to the LHC}
\author{Joachim Kopp$^{1,2}$}            \email[Email: ]{jkopp@fnal.gov}
\author{Ethan T.~Neil$^1$}              \email[Email: ]{eneil@fnal.gov}
\author{Reinard Primulando$^{3}$}  \email[Email: ]{reinard@jhu.edu}
\author{Jure Zupan$^4$}              \email[Email: ]{zupanje@ucmail.uc.edu}
\affiliation{$^1$ Fermilab, P.O.~Box 500, Batavia, IL 60510, USA \\
             $^2$ Max Planck Institut f\"ur Kernphysik, Saupfercheckweg 1, 69117 Heidelberg, Germany \\
             $^3$ Department of Physics and Astronomy, Johns Hopkins University, Baltimore, MD 21218, USA \\
             $^4$ Department of Physics, University of Cincinnati, Cincinnati, Ohio 45221,USA}
\date{January 8, 2013} 
\pacs{}

\begin{abstract}

  We explore the relationship between astrophysical gamma-ray signals and LHC signatures for a class of phenomenologically successful secluded dark matter models, motivated by recent evidence for a $\sim 130$~GeV gamma-ray line.  We consider in detail scenarios in which interactions between the dark sector and the standard model are mediated by a vev-less scalar field $\phi$, transforming as an $N$-plet ($N > 3$) under $SU(2)_L$.  Since some of the component fields of $\phi$ carry large electric charges, loop induced dark matter annihilation to $\gamma \gamma$ and $\gamma Z$ can be enhanced without the need for non-perturbatively large couplings, and without overproduction of continuum gamma-rays from other final states.  We discuss prospects for other experimental tests, including dark matter--nucleon scattering and production of $\phi$ at the LHC, where searches for monophotons, monojets and anomalous charged tracks may be sensitive.  The first LHC hints could come from the Higgs sector, where loop corrections involving $\phi$ lead to significantly modified $h \to \gamma \gamma$ and $h \to \gamma Z$ branching ratios.

\end{abstract}

\begin{flushright}
  FERMILAB-PUB-13-006-T
\end{flushright}

\maketitle

\newpage
\section{Introduction and motivation}
\label{sec:intro}

Recent analyses of Fermi-LAT data have revealed a line-like feature in the cosmic gamma ray energy spectrum from the Galactic Center at an energy $\sim 130$~GeV \cite{Bringmann:2012vr, Weniger:2012tx, Su:2012ft, Fermi:2012}. Additional hints for a 130~GeV photon line were seen in galaxy clusters \cite{Hektor:2012kc} and unassociated Fermi-LAT sources \cite{Su:2012zg} (see, however, \cite{Hooper:2012qc, Mirabal:2012za, Hektor:2012jc}). At present, it is not clear whether these features are due to an instrumental effect or due to physics beyond the Standard Model (SM). Validation tests done in the original Refs.~\cite{Bringmann:2012vr, Weniger:2012tx, Su:2012ft}, as well as additional checks using the public data performed in Refs.~\cite{Bringmann:2012ez, Whiteson:2012hr, Finkbeiner:2012ez, Hektor:2012ev, Rao:2012fh}, have so far not identified an obvious problem with the data, but an official analysis by the Fermi-LAT collaboration will certainly shed further light on the issue.

In this paper we assume that the signal is evidence for dark matter (DM) particles $\chi$ annihilating into two photons, $\chi\chi \to \gamma\gamma$, or a photon and a $Z$ boson, $\chi\chi\to Z\gamma$. In the former case, the DM would need to have a mass $M_\chi\sim130$~GeV and an annihilation cross section  $\ev{\sigma(\chi\chi\to \gamma\gamma) v_{\text{rel}}} \sim 1.3 \times 10^{-27} \text{cm}^3/\text{s}$~\cite{Weniger:2012tx}, whereas if the signal is due to the annihilation process $\chi\chi\to Z\gamma$, one obtains $M_\chi\sim 144$~GeV and  $\ev{\sigma v_{\text{rel}}}\sim 3.1 \times 10^{-27} \text{cm}^3/\text{s}$ \cite{Bringmann:2012ez}.
The fact that we see a photon signal requires that DM couples to a state $\phi$ that is charged under the electroweak gauge group.  Annihilation can then proceed through $\phi$ loops. The required annihilation cross section $\sigma v_{\text{rel}}$ is roughly an order of magnitude smaller than what is required for a thermal relic, but still large for a loop suppressed process. It is, for instance, much bigger than what is expected from a singly charged particle $\phi^+$ running in the loop, unless the coupling of $\phi^+$ to DM is large, close to the perturbativity limit~\cite{Buckley:2012ws}.  Additionally, if the charged particles $\phi^+$ in the loop are lighter than $m_\chi$, the DM can annihilate into them at tree level. These annihilations would contribute significantly to the continuum photon emission from the galactic center due to final state radiation and decays of secondary pions. The resulting annihilation cross sections are typically excluded by strong bounds on the continuum photon emission from the galactic center~\cite{Buchmuller:2012rc, Cohen:2012me, Cholis:2012fb, Hooper:2012sr, Asano:2012zv}. 

Many models have been proposed to circumvent these problems \cite{Dudas:2012pb, Cline:2012nw, Choi:2012ap, Lee:2012bq, Rajaraman:2012db, Acharya:2012dz, Das:2012ys, Kang:2012bq, Weiner:2012cb, Heo:2012dk, Tulin:2012uq, Li:2012jf, Cline:2012bz, Bai:2012qy, Bergstrom:2012bd, Wang:2012ts, Weiner:2012gm, Lee:2012wz, Baek:2012ub, Shakya:2012fj, Li:2012mr,Schmidt2012,Cvetic:2012kj,D'Eramo:2012rr,Li:2012vd,Bernal:2012cd,SchmidtHoberg:2012ip,Farzan:2012kk,Chalons:2012xf,Dissauer:2012xa,Asano:2012zv,Rajaraman:2012fu,Bai:2012yq,Zhang:2012da,Bai:2012nv,Lee:2012ph,Kyae:2012vi,Park:2012xq,Chu:2012qy}. In this paper we focus on a particular set of models that can lead to interesting signals at the LHC. In these ``secluded dark matter" models, DM couples to the visible sector primarily through loops of a new electroweak multiplet $\phi$.  For concreteness we focus on examples where $\phi$ is a scalar with vanishing vacuum expectation value (vev). The salient features of this type of model are
\begin{itemize}
  \item The DM annihilation cross section to photons is enhanced because some states in the mediator multiplet carry large electric charges.

  \item For suppressed DM--Higgs coupling, the continuum photon bounds are avoided because then the dominant annihilation to $W$ and $Z$ bosons is generated at one loop, and to SM fermions only at two loops. The correct relic density is obtained if $\phi$ is somewhat heavier than the DM.

  \item If the mediator $\phi$ couples to the Higgs boson $h$, the branching ratios for the decays $h\to \gamma\gamma$ and $h\to \gamma Z$ are altered.
  If the new particle discovered recently by the ATLAS and CMS collaborations~\cite{ATLASHiggsgamma, CMSHiggsgamma} is indeed a SM-like Higgs boson, the experiments could see these modified branching ratios in future precision measurements.

\item The charged components $\phi^{n\pm}$ of the mediator multiplet $\phi$ are produced at the LHC through their electroweak gauge couplings. Decays to off-shell $W$ bosons lead to multi-lepton final states, however, for large parts of the parameter space the leptons are  so soft that the signal is not observable at the LHC or the Tevatron. The best probe of $\phi^{n\pm}$ production are then final states with a photon and large missing energy (plus possibly other visible particles) because of the large couplings of $\phi^{n\pm}$ to the photon. Moreover, for very small mass splittings, the lifetimes of the $\phi^{n\pm}$ are so long that they can appear as anomalous charged tracks in the inner detectors of ATLAS and CMS.

   \item All 4-scalar couplings are perturbative and continue to be so up to the Planck scale. In particular, the DM--$\phi$ coupling can be relatively small and still lead to a large gamma ray signal because of the large $\phi^{n\pm}$ charges.   For the large $SU(2)$ representations ($N > 3$) considered here, the weak gauge coupling becomes non-perturbative below the Planck scale, see e.g.~\cite{Kopp:2009xt}.  This implies that perturbative grand unification is only possible if the model is embedded into a more complete theory at an intermediate scale. For $N \leq 9$, the embedding (or, alternatively, non-peturbativity of the weak interaction) does not have to occur at scales below several 100~TeV, outside the reach of the LHC.
\end{itemize}

The connection between the 130~GeV gamma ray line and an enhanced $h\to \gamma\gamma$ signal at the LHC has been made also in \cite{Wang:2012ts} for a model with an electroweak triplet mediator. While in \cite{Wang:2012ts} implications for other LHC searches were not elaborated on, we keep the discussion as general as possible and explore also LHC signals aside from the enhanced Higgs to diphoton rate.  We also consider general electroweak multiplets beyond the triplet, but for numerical examples we will use electroweak quintuplets as mediators. We will discuss to what extent electroweak multiplets are constrained by precision Higgs physics, by searches for anomalous charged tracks, and by monojet, monophoton and photon + MET + X searches. In the context of the 130~GeV gamma ray line, LHC final states with a photon and missing energy were also considered in \cite{Lee:2012ph} in the context of models with $Z'$ and axion mediators. Since in these models, the photon is produced as part of the hard process, mono-photon searches are more constraining than in our models, where photons are only produced radiatively.  Finally, independently of the 130~GeV line, the effects of a scalar electroweak quartet on Higgs boson decays to $\gamma \gamma$ and $Z \gamma$ have been considered previously in \cite{Picek:2012ei}.

The paper is organized as follows. In \cref{sec:model} we introduce the class of models we consider in more detail. Section~\ref{sec:gamma} focuses on DM annihilation into photons and on continuum photon emission bounds. In \cref{sec:constrainDM}  we discuss the cosmological history and prospects for DM direct detection.   Section~\ref{sec:constrainEW} deals with existing electroweak precision constraints. The collider phenomenology is discussed in \cref{sec:collider}, and the modifications of the Higgs boson properties in \cref{sec:higgs}. We conclude in \cref{sec:conc}, while calculational details are relegated to the appendices.

\section{Model setup}
\label{sec:model}

\begin{figure}
  \begin{center}
    \includegraphics[width=0.45\textwidth]{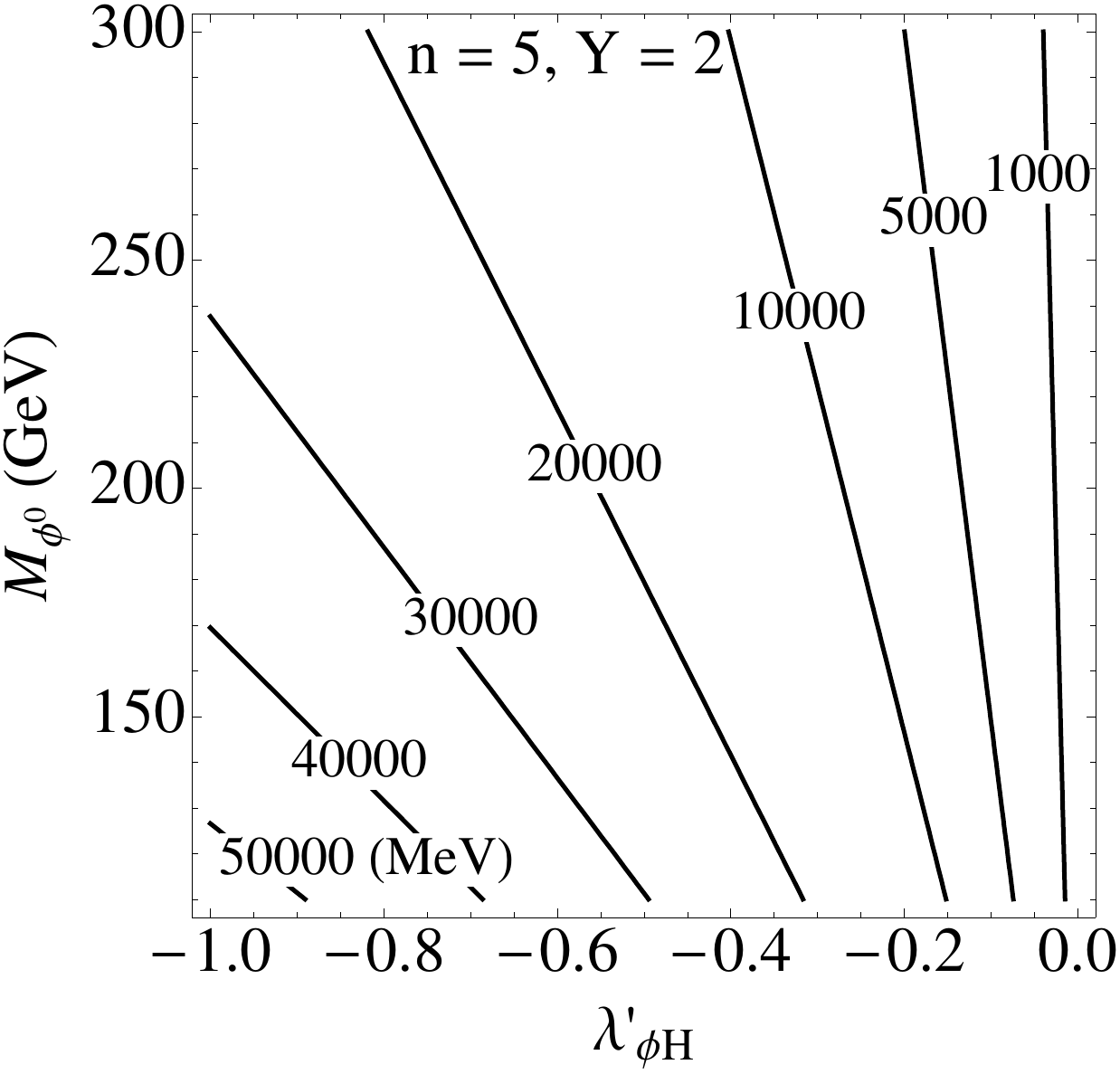}
    \includegraphics[width=0.45\textwidth]{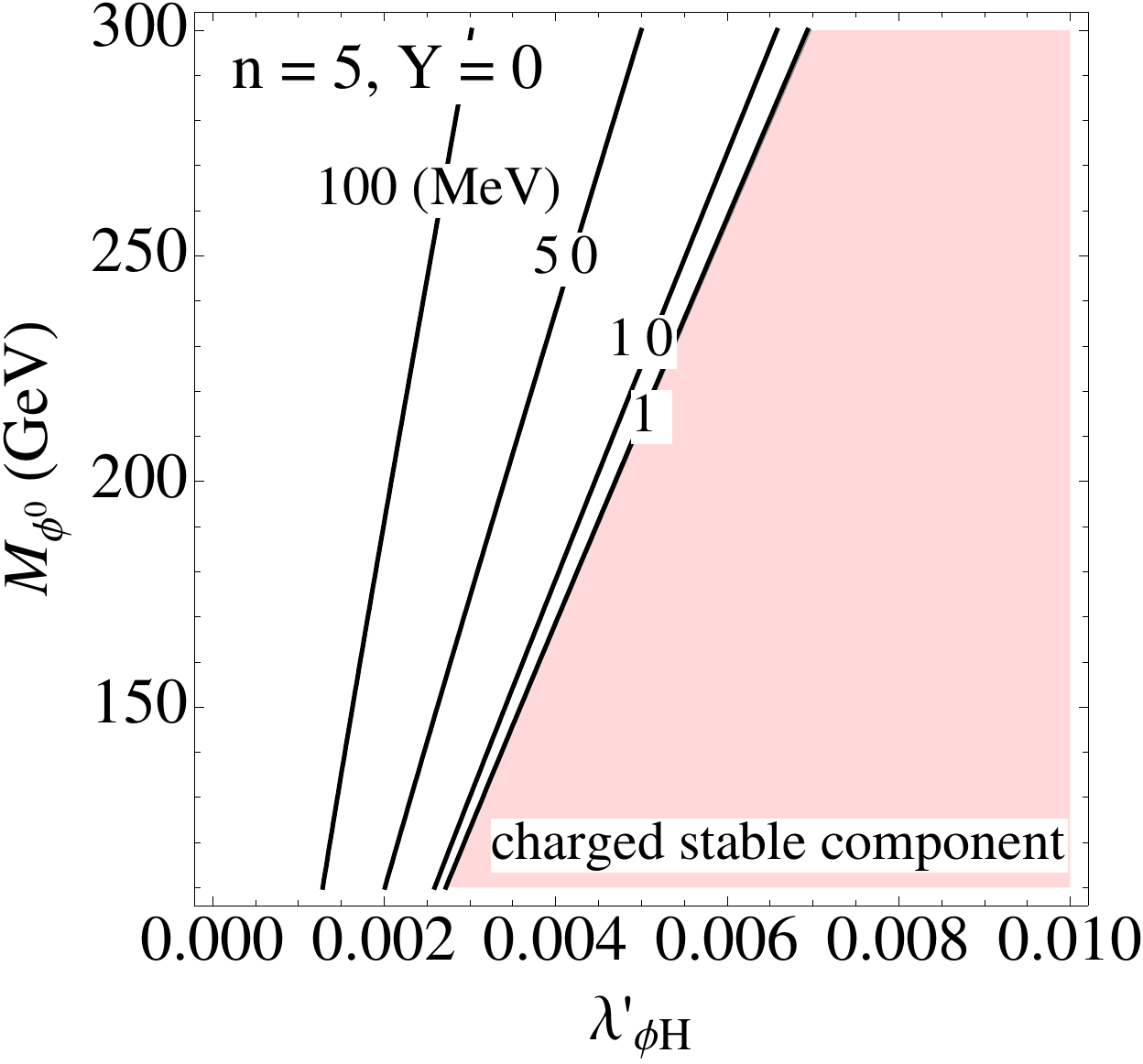}
  \end{center}
  \caption{The mass difference in MeV between the singly charged and neutral mediators, $\phi^+$ and $\phi^0$, as a function of the $\phi^0$ mass $M_{\phi^0}$ and the coupling constant $\lambda^\prime_{\phi H}$ defined in \cref{Lagr-phi}.}
  \label{fig:masssplitting}
\end{figure}

We consider an extension of the SM by a scalar $N$-dimensional $SU(2)_L$ multiplet $\phi$ of hypercharge $Y_\phi$. If 
$N \geq 5$ 
there are no renormalizable couplings to the SM linear in $\phi$.\footnote{This is also true in the case that $N \leq 4$, except for some specific values of $Y_\phi$.}  The $\phi$ fields then interact with the SM only through Higgs portal and gauge interactions,
\beq
\label{Lagr-phi}
\mathcal{L} \supset |D_\mu \phi|^2 - m_\phi^2 \phi^\dagger \phi - \lambda_{\phi H} \phi^\dagger \phi H^\dagger H - \lambda'_{\phi H} (\phi^\dagger T^a_N \phi)(H^\dagger \tau^a H) - \lambda_{4} (\phi^\dagger \phi)^2,
\eeq
where $T^a_N$ and $\tau^a$ are the generators of the $SU(2)_L$  representations $\mathbf{N}$ and $\mathbf{2}$, respectively (their normalization is given in Appendix~\ref{sec:generators}).  We assume the Higgs portal coupling $\lambda_{\phi H}$ to be either positive or negative but with $|\lambda_{\phi H} v^2| \ll m_\phi^2$, so that $\phi$ does not develop a vacuum expectation value. Here, $v$ is the vacuum expectation value (vev) of the Higgs. For the same reason (and other reasons discussed below), $\lambda_{\phi H}'$ should not be too large in magnitude.
Expanding the covariant derivative in \cref{Lagr-phi} gives interactions between $\phi$ and the electroweak gauge fields,
\beq
\begin{split}
\label{Lagr-gauge}
\mathcal{L} \supset& \,\, i\left(\phi_i^\dagger \partial_\mu \phi_j - (\partial_\mu \phi_i^\dagger)\phi_j\right)\left(gA^{\mu,a} (T^a_N)_{ij} + g'Y_{\phi}B^\mu \delta_{ij}\right) \\
&+ \phi_i^\dagger \phi_j \left(\frac{1}{2} g^2 A_\mu^a A^{\mu,b} \big\{T^a_N, T^b_N\big\}_{ij} + g'^2 Y_{\phi}^2 B_\mu B^\mu \delta_{ij} + 2gg' Y_\phi A_\mu^a B^\mu (T^a_N)_{ij}\right).
\end{split}
\eeq
Note that since \cref{Lagr-phi} is the most general renormalizable Lagrangian, the $Z_2$ symmetry $\phi\to -\phi$ is accidental. The neutral component of $\phi$ can thus be stable and a DM candidate in principle.  However, the annihilation process $\phi^0 \phi^0 \rightarrow W^+ W^-$, which occurs at tree level, has too large a cross-section to give the observed DM abundance today for $\phi$ masses below TeV scales \cite{Cirelli:2009uv, Hambye:2009pw}.  Moreover, even with the correct relic density, the same annihilation process for relatively light $\phi$ in the present day would be in tension with observations of dwarf galaxies by the Fermi-LAT collaboration~\cite{Ackermann:2011wa}.  In order to accommodate the tentative Fermi-LAT line, we therefore introduce an additional real vev-less SM-singlet scalar $\chi$, which has direct couplings only to the other scalars:
\beq
\label{Lagr-chi}
\mathcal{L} \supset \frac{1}{2} \partial_\mu \chi \partial^\mu \chi - \frac{1}{2} m_\chi^2 \chi^2 -\lambda_{\chi H} \chi^2 H^\dagger H - \lambda_{\chi \phi} \chi^2 \phi^\dagger \phi.
\eeq
Here, we have introduced by hand a  $Z_2$ symmetry that stabilizes $\chi$, and assumed that $\lambda_{\chi H}$ is chosen such that $\chi$ does not develop a vev. If we wish to explain the tentative Fermi-LAT gamma ray line at $\sim 130$~GeV, the DM mass is fixed at $M_\chi \simeq 130$~GeV or $M_\chi \simeq 144$~GeV, depending on whether decays to $\gamma \gamma$ or $\gamma Z$ are dominant as we discuss below.  In general, however, $m_\chi$ can be arbitrary.  The mass parameter of the weak multiplet, $m_\phi$, is also free, but as we will see below, phenomenologically most interesting is the region where $M_\phi \gtrsim M_\chi$. After electroweak symmetry breaking, the physical masses of the particles in the $\phi$ multiplet, $M_{\phi^{n\pm}}$, receive an additional contribution from the Higgs vev, $v = 246$~GeV.  The coupling $\lambda_{\phi H}$ in \eqref{Lagr-phi} leads to an overall shift, while the $\lambda'_{\phi H}$ term after EWSB,
\beq
-\lambda'_{\phi H}(\phi^\dagger T^a_N \phi)(H^\dagger \tau^a H) \rightarrow
+ \frac{1}{4} \lambda'_{\phi H} v^2 \phi^\dagger T^3_N \phi \,,
\label{eq:lambda-phiHprime}
\eeq
leads to a mass splitting
\beq
\Delta (M^2) = -\frac{1}{4} \lambda'_{\phi H} I_3 v^2,
\label{eq:mass-splitting-lambda-prime}
\eeq
between the $\phi$ component with $T^3_N$ eigenvalue $I_3$ and the $T^3_N = 0$ component. 
There are three interesting regimes of the $\lambda_{\phi H}'$ coupling; for $\lambda_{\phi H}' \sim {\mathcal O}(1)$ the splitting is tens of GeV,  for $\lambda_{\phi H}'\sim {\mathcal O}(0.1)$ the splitting is several GeV, while for $\lambda_{\phi H}'=0$ a splitting arises only from one loop electroweak corrections and is tens to hundreds of MeV as shown in \cref{fig:masssplitting} (right). For large mass splittings the decays of charged $\phi$ particles are easily observable at the LHC and are excluded, so we will be interested in smaller values of $\lambda_{\phi H}'$, of ${\mathcal O}(0.1)$ or below. The parameter $\lambda_{\phi H}'$ in general cannot be made arbitrarily small without fine-tuning since it can be generated from a loop with $A_\mu^a$ and $B_\mu$ on the two internal lines. For $Y_\phi=0$, however, this contribution is zero (cf.~\cref{Lagr-gauge}) so that $\lambda_{\phi H}' \simeq 0$ is natural in this case. For $Y_\phi\ne0$ there is a log divergent contribution to the bare $\lambda'_{\phi H}$ coupling. Even if the cut-off of the theory is at the Planck mass, however, such a contribution is only $\log (M_{Pl} / M_W) 2 Y_\phi \alpha / 4\pi \sim 0.1$,  so that the values of $\lambda_{\phi H}'$ chosen in \cref{fig:masssplitting} (left) are natural. 

As already mentioned, for small values of $\lambda_{\phi H}'$ an important contribution to the mass splitting are the 1-loop electroweak radiative corrections.
The resulting mass splitting is given by~\cite{Cirelli:2009uv}
\beq
\begin{split}
M_Q - M_{Q'} = \frac{\alpha s_W^2 M_\phi}{4\pi}\Big\{ &(Q^2 - Q'^2)s_W^2 f(M_Z/M_\phi) \\ 
&+(Q - Q')(Q + Q' + 2Y_\phi) [f(M_W/M_\phi) - f(M_Z/M_\phi)] \Big\},
\label{eq:mass-splitting}
\end{split}
\eeq
where $Q$ and $Q'$ are the electromagnetic charges of two component fields of $\phi$, $\alpha$ is the fine structure constant, $Y_\phi$ is the hypercharge of the multiplet, and $s_W=\sin\theta_W$ is the sine of the Weinberg angle. The loop function $f$ is given by
\begin{align}
  f(r) &= -r \left[2r^3\ln r+ (r^2-4)^{3/2} \ln A \right]/4,
  \qquad \text{with } A= (r^2 - 2 -r \sqrt{r^2-4})/2 \,.
\end{align}
Here, the UV divergence has been absorbed into the renormalization of $M_\phi$ and $\lambda_{\phi H}'$ (we are using the scheme $k=0$ in the notation of \cite{Cirelli:2009uv}). Numerically, $f(1)=2.72$, so that for $M_\phi \sim {\mathcal O}(100~\text{GeV})$ the mass splitting due to electroweak corrections is tens of MeV.

Because of the accidental $Z_2$ symmetry in \cref{Lagr-phi}, the lightest component of $\phi$ is stable. If we view the model only as a low energy effective theory, however, the Lagrangian in \cref{Lagr-phi} is supplemented by higher dimensional operators which can allow the lightest component of $\phi$ to decay. If $\phi$ forms an $N$-dimensional multiplet, then the lowest dimensional operator mediating this decay needs to contain at least $N-1$ SM doublets. For example, the choice $N=5$, $Y_\phi = 2$ allows us to include the operator
\beq
\mathcal{L}_5 \supset \frac{c_{\phi}}{\Lambda} \phi (H^\dagger)^4 \,.
\label{eq:decay-operator}
\eeq
For general $N$, operators of this type will be suppressed by $1/\Lambda^{N-4}$, where $\Lambda$ is the cut-off scale of the effective theory.

\begin{table}
   \centering
   \begin{ruledtabular}
   \begin{tabular}{l@{\qquad}cc@{\qquad}cc}
                               & \multicolumn{2}{c}{Benchmark model 1: stable $\phi^0$}
                               & \multicolumn{2}{c}{Benchmark model 2: unstable $\phi^0$} \\
     \hline
     Multiplet $SU(2)$ representation & $N$ & 5 & $N$ & 5 \\
     Multiplet hypercharge     & $Y_\phi$ & 0 & $Y_\phi$ & 2 \\
     DM mass                   & $M_\chi$ & 144 GeV & $M_\chi$ & 130 GeV \\ 
     Multiplet mass parameter  & $m_\phi$ & 199.65 GeV & $m_\phi$ & 168.5 GeV \\
     DM--Higgs coupling        & $\lambda_{\chi H}$ & 0 & $\lambda_{\chi H}$ & 0 \\
     DM--multiplet coupling    & $\lambda_{\chi \phi}$ & 0.954& $\lambda_{\chi \phi}$ & 0.493 \\
     $T^3_N$-indep. $\phi-H$ coupling & $\lambda_{\phi H}$ & $- 0.45$ & $\lambda_{\phi H}$ & $- 0.2$ \\
     $T^3_N$-dep. $\phi-H$ couplings& $\lambda'_{\phi H}$ & 0 & $\lambda'_{\phi H}$ & $- 0.1$ \\
     \hline
     Physical multiplet masses &~~$M_{\phi^{\pm\pm}}$~~ & 162.65 GeV & ~~$M_{\phi^{++++}}$~~ & 159.2 GeV \\
                               & $M_{\phi^{\pm}}$ & 162.11 GeV & $M_{\phi^{+++}}$ & 154.4 GeV \\
                               & $M_{\phi^{0}}$ & 161.92 GeV & $M_{\phi^{++}}$ & 149.4 GeV \\        
                               & & & $M_{\phi^{+}}$ & 144.2 GeV \\
                               & & & $M_{\phi^0}$ & 138.9 GeV \\                
     Multiplet relic density   & $\Omega_{\phi_0} h^2$ & $3.6 \times 10^{-4}$ \\
   \end{tabular}
   \end{ruledtabular}
   \caption{The input parameters, resulting mass spectra and relic densities for the two benchmark points: A $Y_\phi=0$\ \ $5$-plet with stable $\phi^0$ (left), and a $Y_\phi = 2$\ \ $5$-plet with $\phi^0$ allowed to decay through higher-dimensional operators (right).  Note that each $\phi^{n\pm}$ state  is associated with an antiparticle $\phi^{\star,n\pm}$ carrying equal but opposite charge.}
   \label{tab:benchmarkpoint}
\end{table}

In the following, we consider in detail two benchmark cases, one in which we assume that the lightest component of the $\phi$ multiplet is stable on cosmological timescales, and one in which it decays rapidly through higher dimensional operators, see \cref{tab:benchmarkpoint}.  In the stable case, the lightest component of $\phi$ contributes to the dark matter relic density at the subdominant level. For instance, for the benchmark model listed in the left part of \cref{tab:benchmarkpoint}, its relic density  is $\Omega_{\phi^0} h^2 =3.6\times 10^{-4}$. It is thus important that the lightest component of $\phi$ is electrically neutral and does not couple to the $Z$---if it did, its scattering cross section on nuclei would be in conflict with direct detection constraints. To avoid couplings to the $Z$, we have to ensure that $\phi^0$ has $T^3_N = 0$, which is only possible for odd multiplet order $N$ and requires $Y_\phi = 0$. We choose ${N} = 5$ for definiteness, but larger multiplets are also viable. To make sure that $\phi^0$ is indeed the lightest component of $\phi$, we also assume that $\lambda_{\phi H}'$ is small enough that the mass splittings among the components of $\phi$ are dominated by electroweak corrections. For $Y_\phi = 0$ these lead to small positive mass shifts for the $T^3 \neq 0$ charged components compared to the neutral one. We also set $\lambda_{\chi H}=0$ so that there is no $\chi \chi\to h\to WW$ annihilation at tree level. The phenomenological consequences of relaxing this assumption will be addressed below

If $\phi$ can decay through higher-dimensional operators, there are much fewer constraints. For example, if the decay is fast enough, all components of $\phi$ could be charged and there is no constraint on which component is the lightest one.  Here, we will nevertheless assume that the lightest component is electrically neutral.
The complete set of model parameters for the two benchmark cases is given in \cref{tab:benchmarkpoint}. In both of them, we focus on ${N}=5$ multiplets, but we will also comment on higher multiplets below.

\section{Gamma-ray annihilation signal}
\label{sec:gamma}

\begin{figure}
  \begin{center}
    \includegraphics[width=0.7\textwidth]{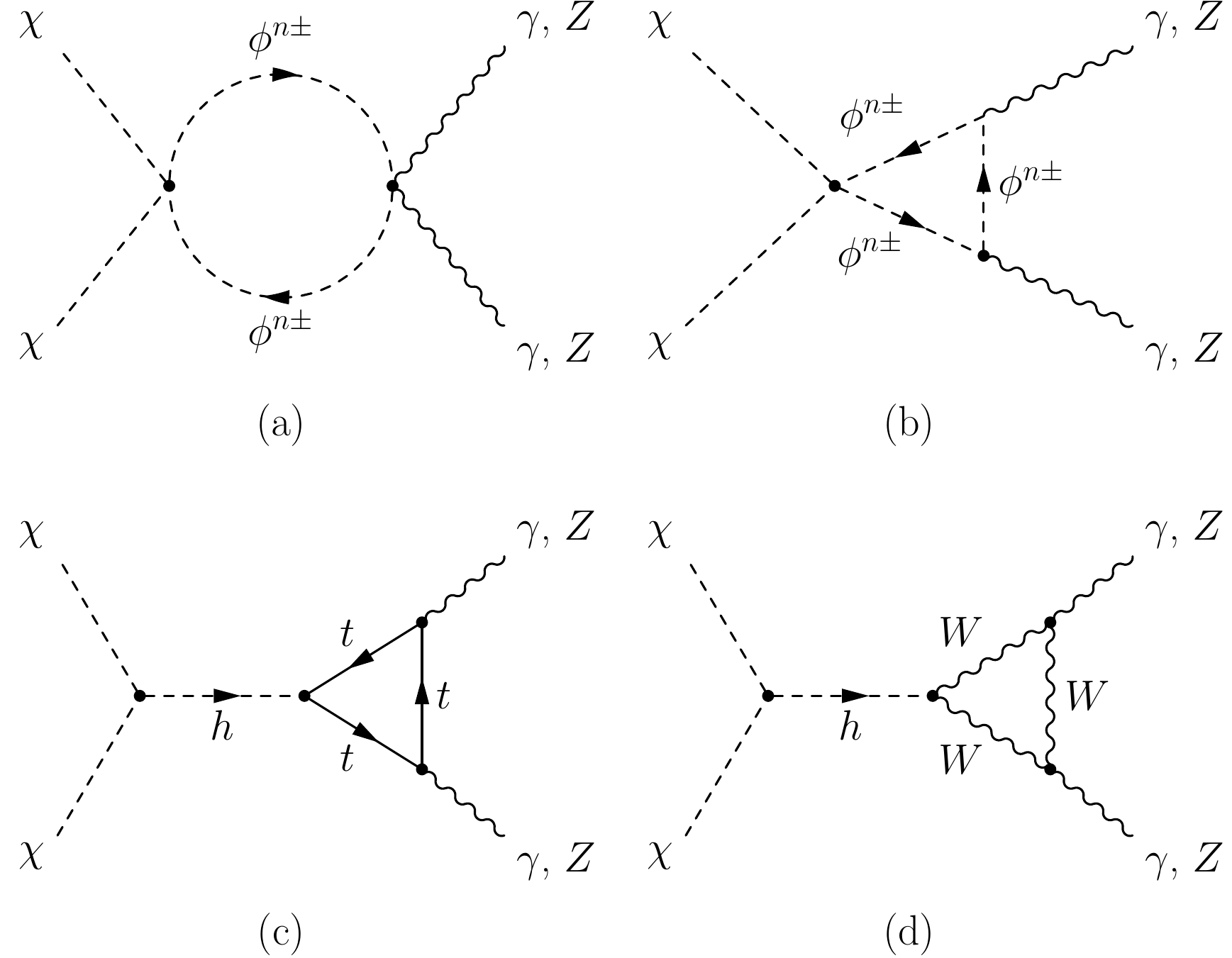}
  \end{center}
  \caption{Representative diagrams contributing to $\chi\chi \to \gamma\gamma$, $\gamma Z$,
  $ZZ$ annihilations}
  \label{fig:chichiAA}
\end{figure}

We are now ready to discuss in detail the phenomenology of the
DM models introduced in \cref{sec:model}, where DM--SM interactions are
mediated by a scalar $SU(2)$ $\mathbf{N}$-plet. We begin by considering
indirect detection constraints, in particular the signals of DM annihilation in the
gamma ray sky. As mentioned in the introduction, one of our motivations is the
tentative line-like feature observed in Fermi-LAT gamma ray data from the
galactic center and other DM-rich regions in the sky~\cite{Bringmann:2012vr,
Weniger:2012tx, Su:2012ft, Fermi:2012, Hektor:2012kc}.  This signal, as well as possible
gamma ray lines that may be discovered in the future, can be due to either
$\chi\chi \to \gamma\gamma$ or $\chi\chi \to \gamma Z$ annihilation. The
process $\chi\chi \to \gamma h$ is not generated in the models we consider
because the initial and final states would have different $C$ parity.  Both
$\chi\chi \to \gamma\gamma$ or $\chi\chi \to \gamma Z$ proceed through diagrams
of the form shown in \cref{fig:chichiAA}. In both of our benchmark points from
\cref{tab:benchmarkpoint} we have $\lambda_{\chi H} = 0$ so that only the topologies
\cref{fig:chichiAA} (a) and (b) contribute. The annihilation cross sections are
then given by~\cite{Shifman:1979eb, Djouadi:2005gi, Djouadi:2005gj}
\begin{align}
  \ev{\sigma v_\text{rel}}_{\gamma\gamma} &= \frac{1}{32 \pi M_\chi^2}
    \bigg|
      \frac{\alpha \lambda_{\chi\phi} }{\pi} \sum Q^2
      \Big(1 - \beta f(\beta) \Big)
    \bigg|^2 \,,
  \label{eq:chichi-gammagamma} \\
\begin{split}
  \ev{\sigma v_\text{rel}}_{\gamma Z} &= \frac{1}{32 \pi M_\chi^2}
    \bigg(1 - \frac{M_Z^2}{4 M_\chi^2} \bigg)^3 \,
    \Bigg|
      \frac{2\sqrt{2} \alpha \lambda_{\chi\phi}}{\pi s_W c_W} \sum Q \big(I_3 - s_W^2 Q\big) \\
  &\quad \times
      \bigg[
          \frac{\gamma}{2(\beta-\gamma)}
        + \frac{\beta \gamma^2}{2(\beta-\gamma)^2} \Big( f(\beta) - f(\gamma) \Big)
        + \frac{\beta \gamma}{(\beta-\gamma)^2}    \Big( g(\beta) - g(\gamma) \Big)
      \bigg]  
    \Bigg|^2,
  \label{eq:chichi-gammaZ}
  \end{split}
\end{align}
with $\beta \equiv M_\phi^2 / M_\chi^2$, $\gamma \equiv 4 M_\phi^2 / M_Z^2$, and
the loop functions 
\begin{align}
  f(x) &= \left\{ \begin{array}{lr}
                    \arcsin^2 \sqrt{x^{-1}}            & \text{for $x \geq 1$}, \\[0.1cm]
                    -\frac{1}{4} \Big[ \log\frac{1 + \sqrt{1-x}}{1 - \sqrt{1-x}} - i \pi \Big]^2 
                                                     & \text{for $x < 1$}.
                  \end{array} \right.
    \label{eq:loop-function-f} \\[0.2cm]
  g(x) &= \left\{ \begin{array}{lr}
                    \frac{\sqrt{1 - x}}{2} \Big[ \log\frac{1+\sqrt{1-x}}{1-\sqrt{1-x}} - i\pi \Big]
                                                     & \text{for $x > 1$}, \\[0.2cm]
                    \sqrt{x-1} \arcsin\sqrt{x^{-1}}  & \text{for $x \leq 1$}.
                  \end{array} \right.
    \label{eq:loop-function-g}
\end{align}
In the above expressions, $s_W$ and $c_W$ denote the sine and the cosine of the
Weinberg angle, respectively, $Q$ and $I_3$ are the electric charge and the third
component of the weak isospin of the $\phi$ field components, and the sums run
over these components.

\begin{figure}
  \begin{center}
    \includegraphics[width=0.48\textwidth]{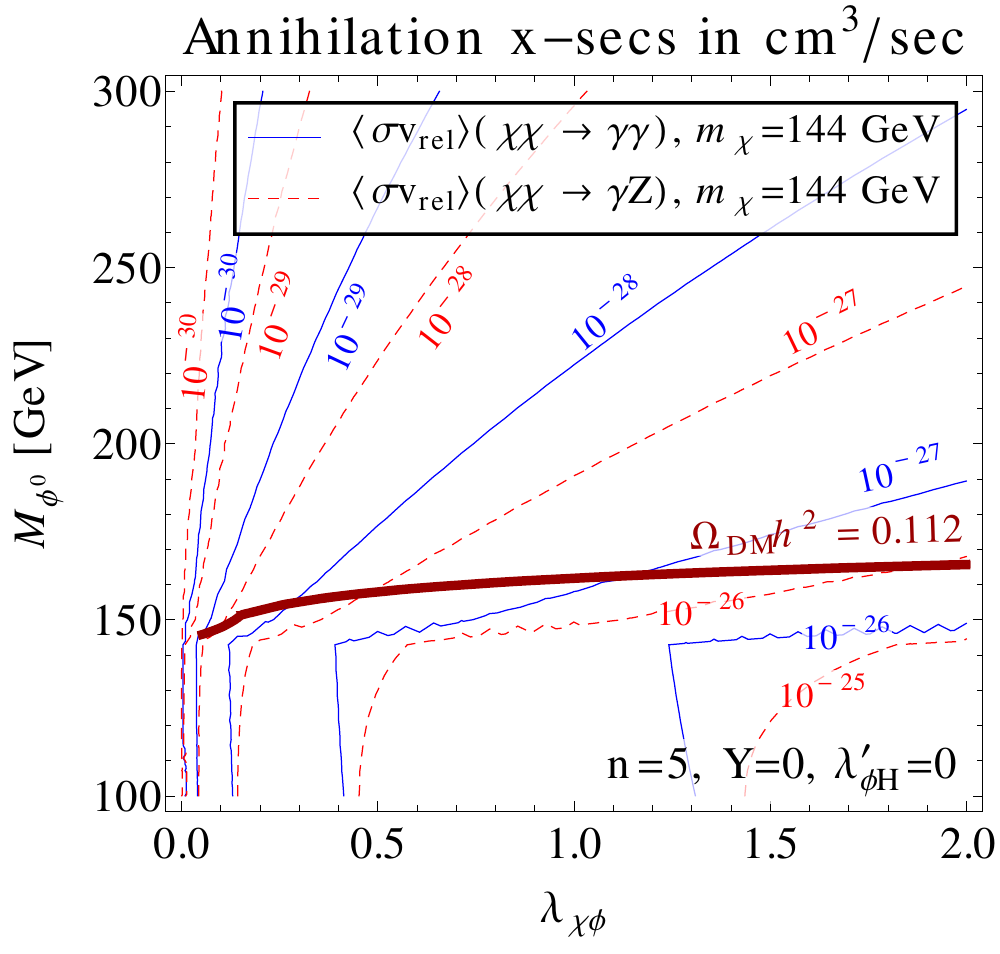}
    \quad\includegraphics[width=0.48\textwidth]{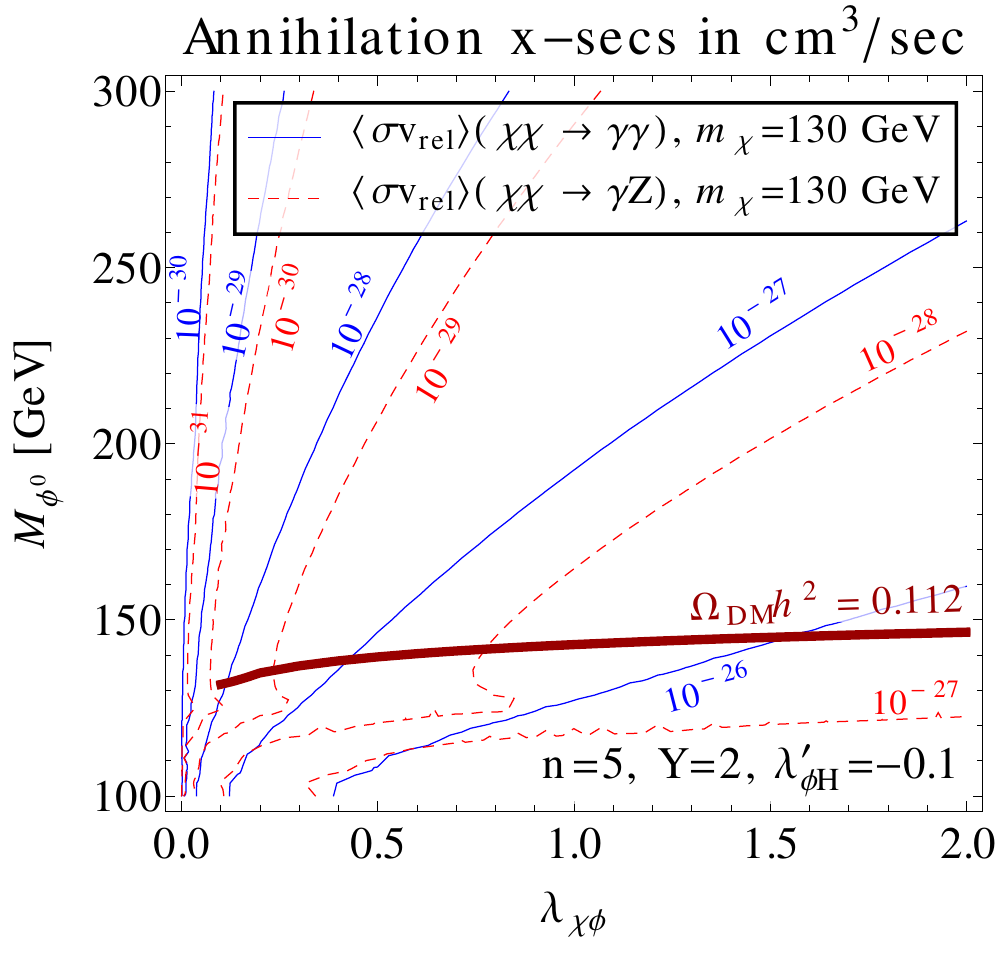}
  \end{center}
  \caption{Contours of constant annihilation cross section $\ev{\sigma
    v_\text{rel}}$ for the annihilation processes $\chi\chi \to \gamma\gamma$
    (blue solid lines) and $\chi\chi \to \gamma Z$ (red dashed lines) as a
    function of $M_{\phi^0}$ (the mass of the neutral component of the mediator
    multiplet $\phi$) and $\lambda_{\chi\phi}$ (the coupling of DM to $\phi$). Motivated by the two benchmark models given in
    \cref{tab:benchmarkpoint}, we take $\phi$ to be an $SU(2)$ 5-plet with
    $Y_\phi = 0$ and no isospin-dependent couplings to the Higgs ($\lambda'_{\phi
    H} = 0$) in the left panel, whereas in the right panel we chose
    $Y_\phi = 2$ and $\lambda'_{\phi H} = -0.1$. Our choice of DM mass,
    $M_\chi=144$~GeV for $\chi\chi \to \gamma Z$ and $M_\chi=130$~GeV for
    $\chi\chi \to \gamma\gamma$ is motivated by the tentative Fermi-LAT gamma ray
    line signal~\cite{Bringmann:2012vr, Weniger:2012tx, Su:2012ft,
    Hektor:2012kc}.  The thick red line denotes the values of $M_{\phi^0}$ and
    $\lambda_{\chi \phi}$ for which the correct DM relic density
    $\Omega_\text{DM} h^2 = 0.112$~\cite{Komatsu:2010fb} is obtained if all
    the other model parameters are fixed 
    as in
    \cref{tab:benchmarkpoint}. The error on the relic density from WMAP, $\pm
    0.0056$, is below the resolution of the plot.}
  \label{fig:sigmav-1}
\end{figure}

The predicted values of the annihilation cross sections $\ev{\sigma
v_\text{rel}}_{\gamma\gamma}$ and $\ev{\sigma v_\text{rel}}_{\gamma Z}$ are
shown in \cref{fig:sigmav-1}.  These should be compared to $\ev{\sigma
v_\text{rel}}_{\gamma\gamma} = (1.27^{+0.37}_{-0.43}) \times
10^{-27}$~cm$^3$/s and \ $\ev{\sigma v_\text{rel}}_{\gamma Z} =
(3.14^{+0.89}_{-0.99}) \times 10^{-27}$~cm$^3$/s, which for the Einasto DM profile were shown
in~\cite{Weniger:2012tx,Bringmann:2012ez} to explain the Fermi-LAT feature at
$130$ GeV for DM masses $m_{\chi}=130$ GeV and $m_{\chi}=144$ GeV, respectively.
We see that annihilation cross
sections $> 10^{-27}$~cm$^3$/s are easily obtained in our model for
$\lambda_{\chi \phi}$ well within the perturbative regime, and without the need
for tuning between $M_{\phi^0}$ and $M_\chi$. Notice the qualitative change in
the dependence of $\ev{\sigma v_\text{rel}}_{\gamma\gamma}$ on $M_{\phi^0}$ and
$\lambda_{\chi\phi}$ when $M_{\phi^0}$ approaches $M_\chi$. The reason is that
for $M_{\phi_i} < M_\chi$, the loop diagrams in \cref{fig:chichiAA} acquire
an imaginary part because direct annihilation $\chi\chi \to \phi\phi$ becomes
possible.  This effect is much more pronounced for the stable benchmark point (\cref{fig:sigmav-1}, left plot) due to the near-degeneracy of the components of $\phi$.  In the unstable case, the non-zero hypercharge assignment allows for destructive interference in $\chi \chi \rightarrow Z \gamma$ for $M_{\phi^0} \sim 125$ GeV, leading to the ``kink" visible in the right plot of \cref{fig:sigmav-1}.

\begin{figure}
  \begin{center}
    \includegraphics[width=0.48\textwidth]{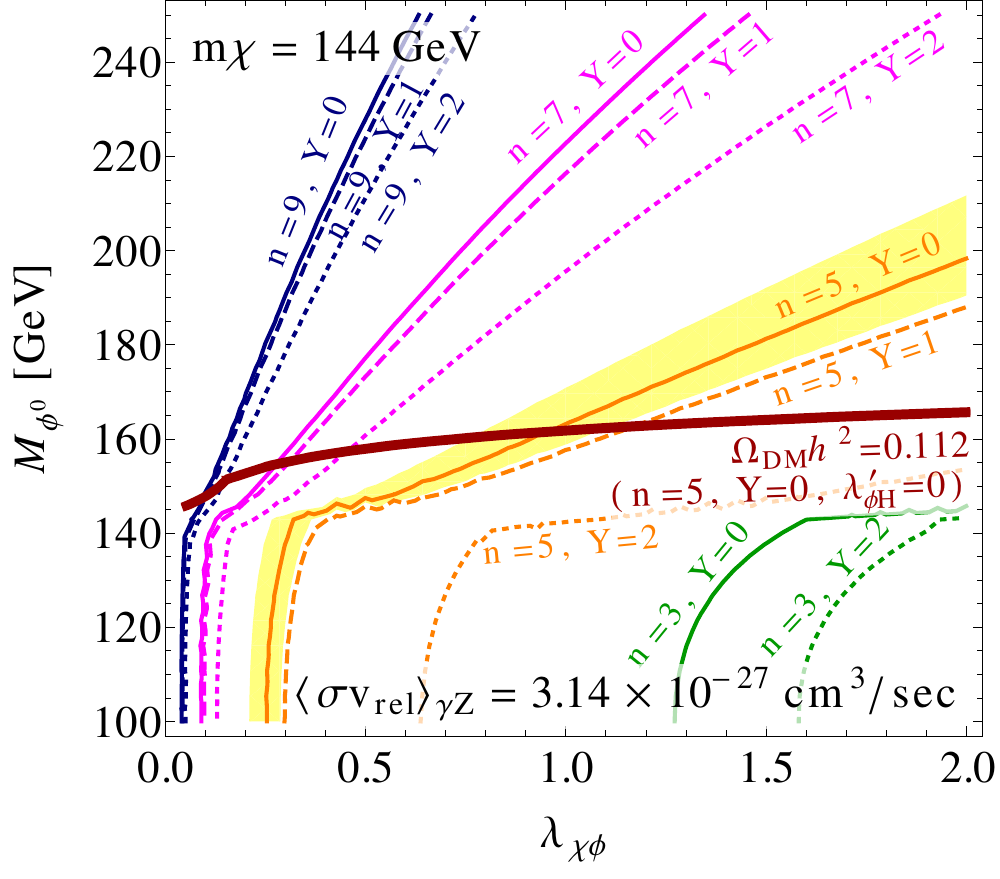} \hfill
    \includegraphics[width=0.48\textwidth]{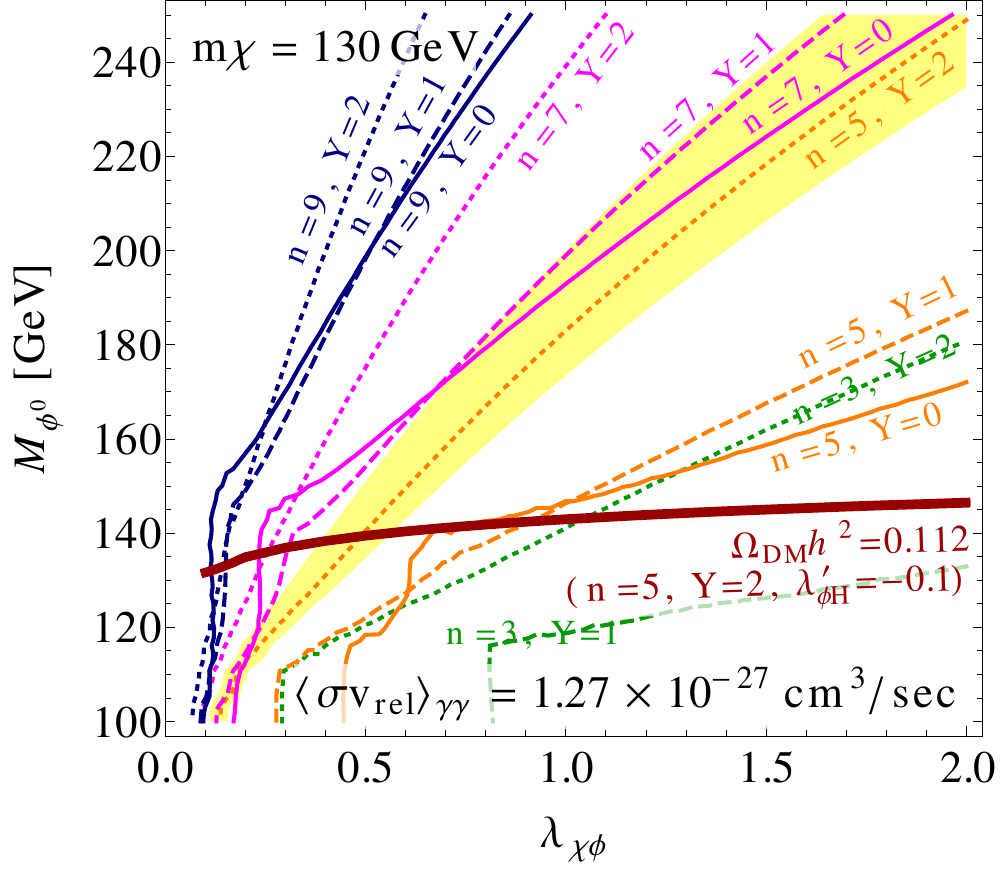}
  \end{center}
  \caption{Contours of constant annihilation cross section 
    $\ev{\sigma v_\text{rel}}_{\gamma Z} = 3.14 \times 10^{-27}$~cm$^3$/s (left) and $\ev{\sigma
    v_\text{rel}}_{\gamma\gamma} = 1.27 \times 10^{-27}$~cm$^3$/s 
    (right), motivated by the tentative Fermi-LAT gamma ray
    line~\cite{Bringmann:2012vr, Weniger:2012tx, Su:2012ft, Hektor:2012kc}, as
    a function of $M_{\phi^0}$ (the mass of the neutral component of the 
    mediator $SU(2)$ multiplet $\phi$) and $\lambda_{\chi\phi}$ (the coupling of DM
    to $\phi$). The results are shown for multiplet sizes $N=3,5,7,9$ (green, orange,
    magenta and blue lines), and hypercharge $Y_\phi=0,1,2$ (solid, dashed, dotted lines).
    The choices of DM mass, $M_\chi=144$ for $\chi\chi \to \gamma Z$ (left panel)
     and $M_\chi=130$~GeV for $\chi\chi \to \gamma\gamma$ (right panel)
     are also motivated
    by the Fermi-LAT line. The remaining input parameters are as in the
    corresponding columns of \cref{tab:benchmarkpoint}.  The yellow bands for
    the benchmark models $N=5$, $Y_\phi=0$ (left) and $N=5$, $Y_\phi=2$ (right)
    show the $1\sigma$ experimental ranges for \ $\ev{\sigma v_\text{rel}}_{\gamma Z} =
    (3.14^{+0.89}_{-0.99}) \times 10^{-27}$~cm$^3$/s and $\ev{\sigma
    v_\text{rel}}_{\gamma\gamma} = (1.27^{+0.37}_{-0.43}) \times
    10^{-27}$~cm$^3$/s, respectively (obtained
    using an Einasto halo profile in~\cite{Weniger:2012tx,Bringmann:2012ez}).  The
    thick red line denotes the values of $M_{\phi^0}$ and $\lambda_{\chi \phi}$
    for which the correct DM relic density $\Omega_\text{DM} h^2 =
    0.112$~\cite{Komatsu:2010fb} is obtained. The error on the relic density
    from WMAP, $\pm 0.0056$, is below the resolution of the plot.}
  \label{fig:sigmav-2}
\end{figure}

To illustrate the dependence of $\ev{\sigma v_\text{rel}}_{\gamma\gamma}$ and
$\ev{\sigma v_\text{rel}}_{\gamma Z}$ on the quantum numbers of $\phi$, we show
in \cref{fig:sigmav-2} left (right) contours of constant $\ev{\sigma
v_\text{rel}}_{\gamma Z \, (\gamma\gamma )} =  3.14 \ (1.27) \times
10^{-27}$~cm$^3$/s, i.e.\ for the central values of annihilation cross
sections motivated by the tentative gamma ray line at
130~GeV~\cite{Weniger:2012tx, Bringmann:2012ez}. Several different choices for
the multiplet dimension $\mathbf{N}$ and its hypercharge $Y_\phi$ are shown. As
expected, it is easiest to obtain the annihilation cross sections required to
explain the 130~GeV line in models with large $\mathbf{N}$ and thus highly
charged component fields of $\phi$.  Note that $\ev{\sigma
v_\text{rel}}_{\gamma\gamma}$ increases with $Y_\phi$ because higher charge
states appear for large $Y_\phi$, whereas $\ev{\sigma v_\text{rel}}_{\gamma Z}$
decreases with $Y_\phi$ because of stronger cancellation between $I_3$ and
$s_W^2 Q$ in the term at the end of the first line of \cref{eq:chichi-gammaZ}.

Besides the annihilation to two photons, the DM in our model also
annihilates to $W^+W^-$ and $ZZ$. If we set the DM-Higgs coupling
$\lambda_{\chi H}$ to zero, as in our benchmark points from
\cref{tab:benchmarkpoint}, annihilations to $W^+W^-$ and $ZZ$ first occurs at 1
loop level. The annihilation cross section is then smaller than the bounds from
continuum gamma rays in Fermi-LAT. Using {\tt FeynArts} \cite{Hahn:2000kx}, we
estimate that for the benchmark point with the stable 5-plet (left part of
\cref{tab:benchmarkpoint}), the annihilation cross section to $W^+W^-$ is $\ev{\sigma v_\text{rel}} = 2.0
\times 10^{-26}\ \text{cm}^3/\text{s}$, and the one to $ZZ$ is $\ev{\sigma v_\text{rel}} = 5.2 \times
10^{-27}\ \text{cm}^3/\text{s}$.
In the case of the unstable 5-plet benchmark point (right part of
\cref{tab:benchmarkpoint}), the annihilation cross section to
$W^+W^-$ is $\ev{\sigma v_\text{rel}} = 5.3 \times 10^{-27}\ \text{cm}^3/\text{s}$ and the one to $ZZ$
is $\ev{\sigma v_\text{rel}} = 2.6 \times 10^{-27}\ \text{cm}^3/\text{s}$. The bound from
continuum gamma rays from the galactic center is $\ev{\sigma v_\text{rel}} = 2.7 \times 10^{-26}
\ \text{cm}^3/\text{s}$ for annihilation to $W^+W^-$ and $\ev{\sigma v_\text{rel}} = 3.2 \times
10^{-26}\ \text{cm}^3/\text{s}$ for the $ZZ$ final
state~\cite{Hooper:2012sr}. The continuum photon constraints can also be
translated into a constraint on $\lambda_{\chi H}$ which is $\lambda_{\chi
H}\lesssim 0.03$.

\section{Relic density and direct detection}
\label{sec:constrainDM}

We now investigate the dynamics of DM freeze-out in the early Universe for the class of models given by the Lagrangians \eqref{Lagr-phi} and \eqref{Lagr-chi}.  At very high temperatures, the DM $\chi$ is kept in thermal equilibrium through two channels: (i) $s$-channel Higgs exchange $\chi \chi \leftrightarrow h \leftrightarrow WW, ZZ$~\cite{Burgess:2000yq} and (ii) direct coupling to the mediator field $\phi$, $\chi\chi \leftrightarrow \phi\phi$.  $\phi$, in turn, is kept in thermal equilibrium with the SM particles through its electroweak interactions.  The amplitude for process (i) is proportional to the coupling constant $\lambda_{\chi H}$, which is constrained by the requirement that secondary gamma rays from DM annihilations in the Galactic Center today should not overshoot the Fermi-LAT constraints on the gamma ray continuum.
Since generating the correct DM relic density $\Omega h^2 = 0.1120 \pm 0.0056$~\cite{Komatsu:2010fb} through $\chi\chi \leftrightarrow h \leftrightarrow WW, ZZ$ alone is only marginally allowed, we will not entertain this possibility here. Instead, we focus on the case where the correct relic density of DM is determined by the ``forbidden'' annihilation channels \cite{Tulin:2012uq, Griest:1990kh, Jackson:2009kg}, $\chi\chi \to \phi\phi$. These channels are not kinematically accessible for nonrelativistic DM since $M_\chi < M_\phi$. Therefore, they do not contribute to DM annihilations today, avoiding indirect detection constraints. In the early universe, however, they can still be effective if $M_\phi$ is not too much larger than $M_\chi$, so that $\chi\chi \to \phi\phi$ is still accessible from the high-energy tails of the thermal DM energy distribution.  Depending on the quantum numbers of $\phi$, we find that the mass gap $M_\phi-M_\chi$ required to explain the observed relic density is several tens of GeV to 100~GeV.  Using {\tt MicrOMEGAs}~\cite{Belanger:2010gh} we estimate that for our $\mathbf{N} = 5$, $Y_\phi = 0$ benchmark point (left part of \cref{tab:benchmarkpoint}), DM freeze-out occurs at $M_\chi / T \sim 25$. The components of the multiplet $\phi$ remain in thermal equilibrium until $M_\chi / T \sim 26$ (which is equivalent to $M_\phi / T \sim 33$) and have lower relic density than $\chi$. Around freeze-out the DM velocity is $\sim 0.3 \, c$ so that it can still annihilate into the slightly heavier $\phi$. This maintains equilibrium with the thermal bath and as a result, the DM freezes out at around the same time as it would in the simple thermal WIMP scenario, giving the correct thermal relic density. 

From the thick red lines in \cref{fig:sigmav-1,fig:sigmav-2} we see that, for both the stable and unstable $5$-plet benchmark models, the correct relic density can be obtained for a number of different parameter choices. The calculations were performed using {\tt MicrOMEGAs}~\cite{Belanger:2010gh}, but cross-checked by solving the relevant Boltzmann equations numerically in {\tt Mathematica}. In the case of the stable $\phi$, the neutral component $\phi^0$ constitutes part of the DM relic density, but as mentioned in \cref{sec:model}, its abundance is expected to be small because of its efficient annihilation to $W^+ W^-$ and the resulting late freeze-out. Indeed, using {\tt MicrOMEGAs}, we find $\Omega_{\phi^0} h^2 = 3.6 \times 10^{-4}$, which is three orders of magnitude lower that the total DM relic density.

We next discuss direct detection signals in the models we are focusing on. DM interactions with nuclei are described by the following effective operators:
\begin{equation}
\label{ops-basis-EFT}
{\cal L}_\text{eff}=\frac{C_F}{M_\phi^2}\chi^2 F_{\mu \nu}F^{\mu\nu} +\sum_q \frac{C_q}{M_\phi^2} M_q \chi^2 \bar q q,
\end{equation}
where in the interactions with quarks we have already included the required quark mass suppression due to a chirality flip.  For the suppression scale, we have chosen the $\phi$ mass for later convenience, while $C_F$ and $C_q$ are dimensionless Wilson coefficients. 
The effective cross section {\it per nucleon} for DM scattering on a nucleus $(A,Z)$ is~\cite{Weiner:2012cb}
\beq 
\label{xsec-DM-nucleon}
\sigma_N^{SI}\simeq \frac{1}{\pi}\frac{m_p^2}{M_\chi^2}\frac{1}{A^2}\bigg(\alpha Z^2 Q_0 {\cal F}(0) \frac{C_F}{2 M_\phi^2}+ {A \sum_q m_{p}\overline{ f_q^{n,p}} \frac{C_q}{M_\phi^2}} \bigg)^2, 
\eeq
where $A$, $Z$ are the atomic mass and atomic number of the nucleus, respectively, the nuclear coherence scale is $Q_0=\sqrt{6}(0.3 + 0.89A^{1/3})^{-1}\ \text{fm}^{-1}$ \cite{Weiner:2012cb}, and we neglected the momentum dependence in the electromagnetic form factor ${\cal F}(|\vec q|)$, replacing it with  ${\cal F}(0)=2/\sqrt{\pi}$. The average for the matrix elements $f^{n,p}_q$ of the scalar operators in the second term is over neutrons and protons in the nucleus, where we use the values  given in \cite{Belanger:2008sj} and take proton and neutron masses equal. In our model $C_F$ arises at 1-loop from $\phi$ running in the loop, while $C_q$ arise at two loops involving $\phi$ and $Z,W$ exchanges. Numerically, the typical size of the scattering cross section on Xe is 
 \beq
 \sigma_N^{SI}\simeq 5.9 \cdot 10^{-48}\ \text{cm}^2 \bigg(\frac{160\ \text{GeV}}{M_\phi}\bigg)^4
 \bigg(\frac{140\ \text{GeV}}{M_\chi}\bigg)^2 \bigg(\frac{C_F}{1/16\pi^2}+2.3 \frac{C_q}{1/(16\pi^2)^2}\bigg)^2,
 \eeq
where for simplicity we have assumed that $C_q$ is independent of quark flavor. This is not entirely correct in our models, where we have $C_t \sim C_{u,c}=1.9 \times 10^{-5}$ and $C_{d,s,b}=2.1\times 10^{-5}$ for stable and unstable 5-plet benchmark points, 
respectively. (Here we have evaluated the two loop integrals in the limit $M_\phi\gg M_{W,Z}$.) For the diphoton operator the numerical values of the Wilson coefficients are $C_F\sim 10^{-3}$. The analytical expressions are  given in Appendix~\ref{sec:app:direct detection}. The numerical values should be compared with the present XENON100 bound, which for $\sim 140$ GeV DM is $\sigma_N^{SI}\lesssim 4\cdot 10^{-45}\ \text{cm}^2$~\cite{Aprile:2012nq}.

Note that  the $C_q$ Wilson coefficient also receives a nonzero tree level contribution due to single Higgs exchange, 
$C_q^\text{tree}=-\lambda_{\chi H} {M_\phi^2}/{M_H^2}$. The coupling $\lambda_{\chi H}$ is bounded from the continuum photon flux to be $\lambda_{\chi H}\lesssim 0.03$ (see Appendix~\ref{sec:treelevel}), and from direct detection it is bounded to be also below $\lambda_{\chi H}\lesssim 0.03$. In our benchmark points we set $\lambda_{\chi H}=0$.

\section{Precision electroweak constraints \label{sec:constrainEW}}

With the addition of a new charged multiplet at scales not far above the electroweak-breaking scale, we might worry that there will be severe constraints from current experimental bounds on precision electroweak observables.  Because our new multiplet does not couple directly to any SM fermions or induce any new symmetry breaking, we expect no significant corrections to flavor physics observables or processes such as anomalous electric dipole moments.  We will therefore restrict our attention to observables linked closely to the gauge sector, namely the running of the gauge couplings, the $S$, $T$, and $U$ parameters, and contributions to the anomalous magnetic moment $(g-2)$ of the electron and muon.

The presence of new charged matter, particularly in a high representation of the gauge group, can have a significant impact on the running of the gauge couplings $g'$ and $g$ (corresponding to $U(1)_Y$ and $SU(2)_L$, respectively.)  If the rate of running is increased substantially, the gauge couplings can become non-perturbative at relatively low energy scales.  
We will not insist on perturbativity of the couplings all the way up to the Planck mass, but only up to multi-TeV cutoff scales, ensuring that our theory is valid at LHC-accessible energy scales.

The contribution of the multiplet $\phi$ to the one-loop $\beta$-function coefficient $b_i$ is given by \cite{Kopp:2009xt}
\beq
b_{\phi, i} = \frac{1}{3} T_i(\mathbf{N}),
\eeq
where for $U(1)_Y$ the quantity $T_i(\mathbf{N}) = Y_\phi^2$, while for $SU(2)_L$ it is equal to the trace invariant $C(\mathbf{N}) = (N^3 - N) / 12$.  For our stable benchmark case $Y_\phi=0$, while in the unstable benchmark it is set to $Y_\phi = (N-1)/2$.  In either case, the more stringent constraint on perturbativity comes from the $SU(2)_L$ running, due to the stronger scaling of $b_{\phi, i}$ with $N$.  For the choice ${N} = {5}$, we have $b_{\phi, i} = 10/3$, and the gauge coupling remains perturbative up to the Planck scale \cite{Kopp:2009xt}; higher representations would give stronger running, with ${N} = {9}$ leading to a breakdown in perturbativity at a scale on the order of $1000$~TeV.

We turn now to the ``oblique parameters" $S$, $T$ and $U$ \cite{Peskin:1991sw}, which encapsulate generic contributions of new electroweak-charged physics objects to low-energy observables.  These parameters are related to the transverse vacuum polarization of electroweak gauge bosons $\Pi_{AB}(p)$. From \cite{Peskin:1991sw}, eq.~(3.12), the parameters are given by
\begin{align}
\alpha S &\equiv 4e^2 [\Pi_{33}'(0) - \Pi_{3Q}'(0)],\\
\alpha T &\equiv \frac{e^2}{s^2c^2 M_Z^2} [\Pi_{11}(0) - \Pi_{33}(0)],\\
\alpha U &\equiv 4e^2 [\Pi_{11}'(0) - \Pi_{33}'(0)].
\end{align}
Considering $S$ first, for a scalar particle and using the definition $Q = T^3 + Y$, the polarization amplitudes can be split diagram by diagram:
\beq
S \propto \Pi_{33}'(0) - (\Pi_{33}'(0) + \Pi_{3Y}'(0)) \rightarrow -\Pi_{3Y}'(0).
\eeq
If there is no significant mass splitting within the multiplet, then this amplitude is proportional (at leading order) to $\tr(T^3)$, which vanishes identically in any representation.  However, in the presence of such a mass splitting, there will be in general a non-zero contribution to $S$.  Computing the relevant one-loop amplitude, we find that
\beq
S = -\frac{4 \alpha Y_\phi}{3 \sin \theta_w \cos \theta_w} \sum_{I_3>0} \left[ I_3 \log \left(\frac{M_{\phi,Y_\phi+I_3}^2}{M_{\phi,Y_\phi-I_3}^2}\right) \right],
\eeq
where the sum is over all positive $I_3$ values for the multiplet $\phi$, and $M_{\phi,Q}$ denotes the mass of the $\phi$ state with electric charge $Q$.  Details of the calculation are shown in appendix~\ref{sec:Sparam}.  For the unstable $\phi$ $5-$plet benchmark 
point given in \cref{tab:benchmarkpoint}, we find the contribution $S \approx -0.016$.  This is not large enough to cause any tension with the experimental constraint $S = 0.00^{+0.11}_{-0.10}$~\cite{Beringer:1900zz}, but in principle a larger multiplet with a substantial mass splitting could be constrained by $S$.

The calculation of the $T$ parameter is similar, but slightly more involved; details are given in appendix~\ref{sec:Sparam}.  We find
\begin{align}
T &= \frac{-\alpha}{2 \sin^4 \theta_w M_W^2} \sum_{s,s'} \left[  \frac{1}{4} s(N-s) \delta_{s+1,s'} + \frac{1}{4}(s-1)(N-s+1) \delta_{s-1,s'} \right] \nonumber \\
&\times \left[ (M_s^2 + M_{s'}^2) - \frac{2M_s^4}{M_s^2 - M_{s'}^2} \log (M_s^2 / M_{s'}^2) \right].
\end{align}
Evaluating this expression numerically for our benchmark points yields $T \approx -0.013$ for the stable case, and $T \approx 0.0062$ for the unstable case, both well within the current experimental bounds
$T = 0.02^{+0.11}_{-0.12}$~\cite{Beringer:1900zz}.

Because there is no direct coupling to standard-model fermions, contributions of the new multiplet $\phi$ to the anomalous magnetic moment $(g_\ell - 2)$ of the charged leptons ($\ell = e, \mu, \tau$) will appear starting at two-loop order, through modifications of electroweak vacuum polarization and vertex functions.  Because the new sector does not induce any additional breaking of gauge symmetry, in the limit of $M_\phi \rightarrow \infty$ we expect that all such corrections must vanish due to gauge invariance.  The leading contribution to $(g_\ell - 2)$ should thus scale as $1/M_\phi^2$.  Naive dimensional analysis then gives us the rough estimate
\beq
\Delta a_\mu \equiv (\Delta g_\mu - 2)/2 \approx \frac{g^4}{2(16\pi^2)^2} \frac{M_\mu^2}{M_\phi^2} = \frac{\alpha^2}{32\pi^2} \frac{M_\mu^2}{M_\phi^2},
\eeq
which for $M_\phi \gtrsim 100$ GeV gives a contribution of about $a_\mu \lesssim 2 \times 10^{-13}$, three orders of magnitude below the current experimental uncertainty in $a_\mu$~\cite{Bennett:2006fi,Jegerlehner:2009ry}.  The expected deviation of the electron $(g-2)$ from its SM value is even further from being experimentally constrained.  However, the above estimates do not include a prefactor due to the charges and multiplicity of the $\phi$ components, which could easily be $O(10^2)$ or larger depending on the exact choice of representation and hypercharge.  Although we will not attempt it here, a precise two-loop calculation of the contribution from this new multiplet to $a_\mu$ would be interesting, and could potentially yield a contribution large enough to explain the current discrepancy between theory and experiment in this quantity \cite{Jegerlehner:2009ry}.

\section{Collider phenomenology}
\label{sec:collider}

The only direct coupling between the dark matter $\chi$ and the SM particles in
the class of models discussed here is through the Higgs portal operator
$\chi\chi H^\dag H$, see \cref{Lagr-chi}. However, we have seen at the end of
\cref{sec:constrainDM} that the corresponding coupling constant should be small
in order to avoid constraints from direct detection and from continuum photon
emission in DM annihilation. Therefore, we do not expect the DM production
cross section at the LHC to be large enough to be discovered anytime in the
near future. On the other hand, the components of the mediator multiplet $\phi$
can be produced abundantly at the LHC through their large electroweak couplings.
Their decay phenomenology will depend crucially on the mass splittings between
them, and since these mass splittings can be quite small, very interesting
collider signatures are expected. We will now discuss the collider phenomenology
of the new electroweak multiplet $\phi$ in more detail.

\subsection{$\phi^{n\pm}$ production and decay at the LHC}

\begin{figure}
  \begin{center}
    \includegraphics[width=0.8\textwidth]{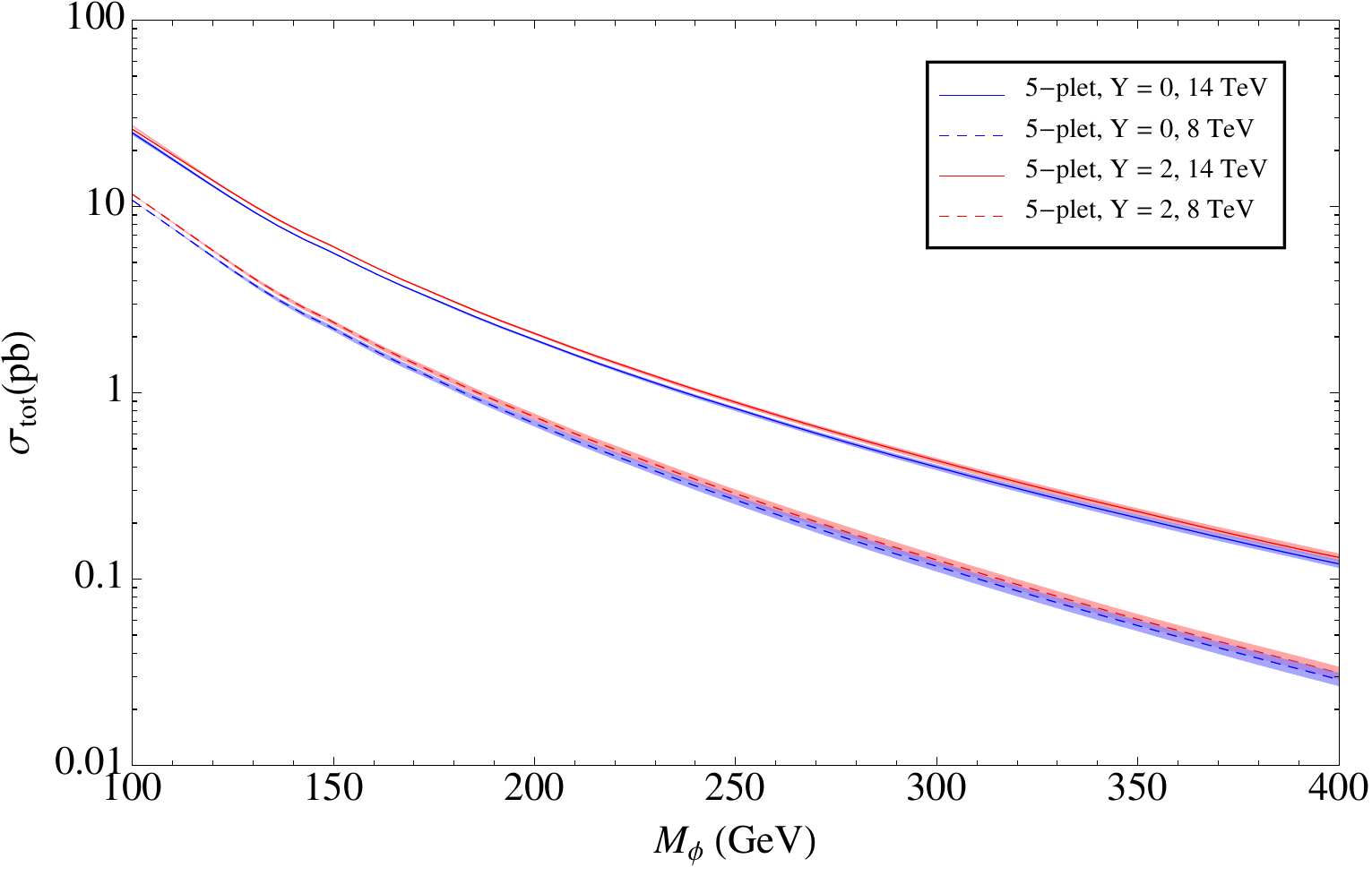}
  \end{center}
  \caption{The LHC pair production cross section for the new electroweak multiplet
    $\phi$ at our benchmark points.
    The blue and red curves are for the $Y_\phi = 0$ and $Y_\phi = 2$ benchmark
    points, respectively. Solid curves are for $\sqrt{s} = 14$~TeV, dashed ones are for
    the 8~TeV LHC. The width of the colored bands indicates the
    theoretical uncertainty of our predictions, estimated by varying the
    factorization and renormalization scales in MadGraph by a factor of 2.}
  \label{fig:crosssection}
\end{figure}

At the LHC, the electroweak multiplet $\phi$ would be produced mostly in
Drell-Yan pair production processes. In \cref{fig:crosssection}, we show the
expected production cross sections at center of mass energies of 8~TeV and
14~TeV.
We see that, especially for relatively light $\phi$ ($M_\phi < 200$~GeV), as at
our benchmark points, the production cross section is fairly large, on the
order of 1~pb at $\sqrt{s} = 8$~TeV and up to more than 10~pb at $\sqrt{s}
= 14$~TeV. Nevertheless, detecting $\phi$ is challenging because by assumption
its lightest component is typically electrical neutral, and the decays of the
heavier components are very soft.

In particular, the charged components of $\phi$ decay via $\phi^{n\pm} \to
\phi^{(n-1)\pm} + W^*$, where the off-shell $W^*$ gives leptons or
hadrons in the final state.  The relevant decay rates are~\cite{Cirelli:2009uv}
\begin{align}
  \Gamma(\phi^{n\pm} \to \phi^{(n-1)\pm} + \pi^\pm) &\simeq
    \frac{(N^2 - 1) V_{ud}^2 f_\pi^2 G_F^2 (\Delta M)^3}{4 \pi}
    \sqrt{1 - \frac{M_\pi^2}{(\Delta M)^2}} \,,
  \label{eq:phi-decay-pi} \\
  \Gamma(\phi^{n\pm} \to \phi^{(n-1)\pm} + e^\pm \parenbar{\nu}_e) &\simeq
    \frac{(N^2 - 1) G_F^2 (\Delta M)^5}{60 \pi^3} \,.
  \label{eq:phi-decay-e}
\end{align}
In these expressions $N$ is the dimension of the $SU(2)$ representation of
$\phi$,  $n = 0 \ldots (N-1)/2$ labels the components of $\phi$, $V_{ud}$ is
the CKM matrix element, $f_\pi \sim 130$~MeV the pion decay constant, $M_\pi$
the pion mass, and $\Delta M$ the mass difference between $\phi^{n\pm}$ and
$\phi^{(n-1)\pm}$. We have made the approximation that $M_\phi \gg \Delta M$,
$M_\pi$. In \cref{eq:phi-decay-e}, we also set the electron mass to zero. In
the case of the $\phi^{n\pm}$ decay to a muon and a neutrino, $\phi^{n\pm}\to \phi^{(n-1)\pm}
\mu^\pm \parenbar{\nu}_\mu$, a similar approximation, $m_\mu\to0$, is not
appropriate, and the analytic expression for the corresponding decay rate is
lengthy. We instead used {\tt CalcHEP}~\cite{Belyaev:2012qa} to compute
the decay rate $\Gamma$ numerically.

It is clear that for small mass splittings $\Delta M$ the hadrons or leptons
produced in $\phi^{n\pm}$ decays are very soft and are thus undetectable in the
LHC's high energy, high luminosity environment.  For instance, for $M_\phi \sim
150$~GeV, even relatively large mass splittings of 5~GeV lead to a lepton with
$p_T > 10$~GeV in only about 2.6\% (3.2\%) of $\phi$ pair production events at
the 8~TeV (14~TeV) LHC. A jet with $p_T > 25$~GeV is produced in only 4.4\%
(5.9\%) of the events.  In these percentages, jets from initial or final state
radiation are not included.

If other energetic final-state particles are present, the cascade decay
products can be boosted and therefore easier to detect.  For example, requiring
a final-state photon with $p_T \geq 80$~GeV leads to leptons with $p_T >
10$~GeV in 7.5\% of $\phi$ pairs produced at the 14~TeV LHC.  However, the
production cross section is reduced to 20~fb.  Existing searches for e.g.~$W +
\gamma$ + MET \cite{ATLAS-WA} are therefore not constraining, and even future
searches in this channel would be challenging, although a search strategy with
more sophisticated kinematic cuts may be more sensitive.

If $\Delta M$ is smaller than $M_\pi$ the hadronic decay modes are
kinematically forbidden and only leptonic modes are allowed. If the splitting
is smaller than the muon mass, only leptonic decays with electrons in the
final state are allowed.  We see, however, from \cref{tab:benchmarkpoint},
that this situation is not realized for our benchmark points.

\subsection{Charged tracks}

\begin{figure}
  \begin{center}
    \includegraphics[width=0.5\textwidth]{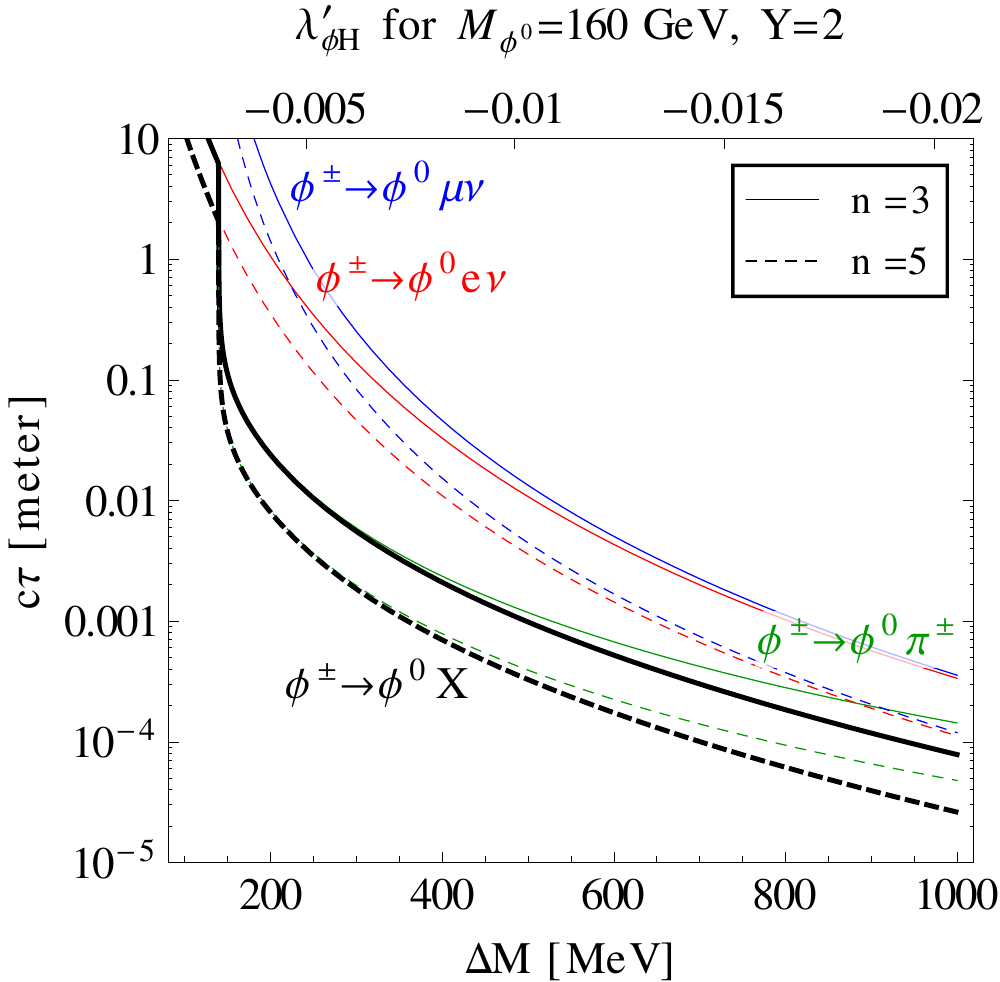}
  \end{center}
  \caption{Lifetime of the singly charged component of $\phi$ as a function
    of the mass splitting $\Delta M$ between $\phi^\pm$ and $\phi^0$.
    The thick black curves show the physical lifetime, taking into account
    hadronic and leptonic decays. The colored curves show the inverse of
    the partial widths to pions (green), $e \nu_e$ (red) and $\mu \nu_\mu$
    (blue). The labels on the upper horizontal axis show the values of
    $\lambda'_{\phi H}$ corresponding to the mass splittings indicated on
    the lower horizontal axis, \emph{neglecting electroweak corrections}.}
  \label{fig:lifetime}
\end{figure}

For very small mass splittings between the components of $\phi$, the $\phi^{n\pm}$ can
travel over macroscopic distances before decaying. This is illustrated
in \cref{fig:lifetime}, which shows that mass splittings below $M_\pi$
are needed for the lifetime of $\phi^{\pm}$ to become macroscopic.
Note that the same relationship between $\Delta M$ and the lifetime $c\tau$
applies also to the multiply charged components of $\phi$. However, in
our benchmark models, the smallest mass splitting and thus the largest
$c\tau$ is always the one corresponding to $\phi^\pm$.

If $\phi^\pm$ decays after travelling more than a few tens of centimeters,
$\phi$ production can be potentially seen in searches for anomalous charged
tracks in the inner detectors of ATLAS and CMS. Since in our benchmark
scenarios, $\phi^\pm$ is part of all $\phi^{n\pm}$ decay chains, such searches
would be sensitive to the production of \emph{any} charged component of $\phi$.
The cross section for this is very similar to the \emph{total} $\phi$ pair
production cross section shown in \cref{fig:crosssection}: Events with
\emph{no} charged $\phi$ (i.e.\ only $\phi^0$) are almost completely absent in
our $Y=0$ benchmark model, where $\phi^0$ does not couple to the $Z$, and they
contribute only about 17\% of the total cross section for the $Y=2$ model.

Searches for charged track signatures
have been carried out by both ATLAS and CMS. We expect the best
sensitivity to our benchmark models to come from future searches of the type
presented by ATLAS in~\cite{ATLAS:2012jp}. In this analysis, a high $p_T$ jet
as well as more than 90~GeV of missing transverse energy are required in
addition to the new charged particle. The latter is required to leave a signal
only in the inner detector, making this search the most sensitive to particles
with lifetimes on the order of several tens of centimeters.

The CMS search for long-lived charged particles~\cite{CMS-PAS-EXO-11-090}
requires signals in the tracking detectors as well as the muon chambers,
implying sensitivity only to particles with decay lengths of order 10~m.
Similarly, the ATLAS search~\cite{Aad:2011mb} requires signals in the inner
detector and the electromagnetic calorimeter. Moreover, only particles with
electric charges $> 6e$ are constrained in this analysis.  The searches from
\cite{CMS-PAS-EXO-11-090} and \cite{Aad:2011mb} are therefore not sensitive to
our benchmark models or minor variations therefore, except for an extremely
fine-tuned corner of parameter space, where electroweak contributions to the
mass splittings, \cref{eq:mass-splitting}, and those induced by nonzero
$\lambda'_{\phi H}$, \cref{eq:mass-splitting-lambda-prime}, conspire to make
one of the mass splittings extremely small.  If we depart further from our
benchmark models, however, it is quite easy to obtain very long-lived charged
particles.  In particular, this is the case if the hypercharge $Y_\phi$ is
chosen such that the lightest component of $\phi$ is charged and decays
only via higher-dimensional operators, for instance
\cref{eq:decay-operator}. Then, its decay width is naturally very small. Note that in
scenarios of this type, the long-lived charged particle should still decay on
timescales $\ll 1$~minute to avoid perturbing big bang nucleosynthesis.

\subsection{Monophoton and monojet signatures}

\begin{figure}
  \begin{center}
    \includegraphics[width=0.8\textwidth]{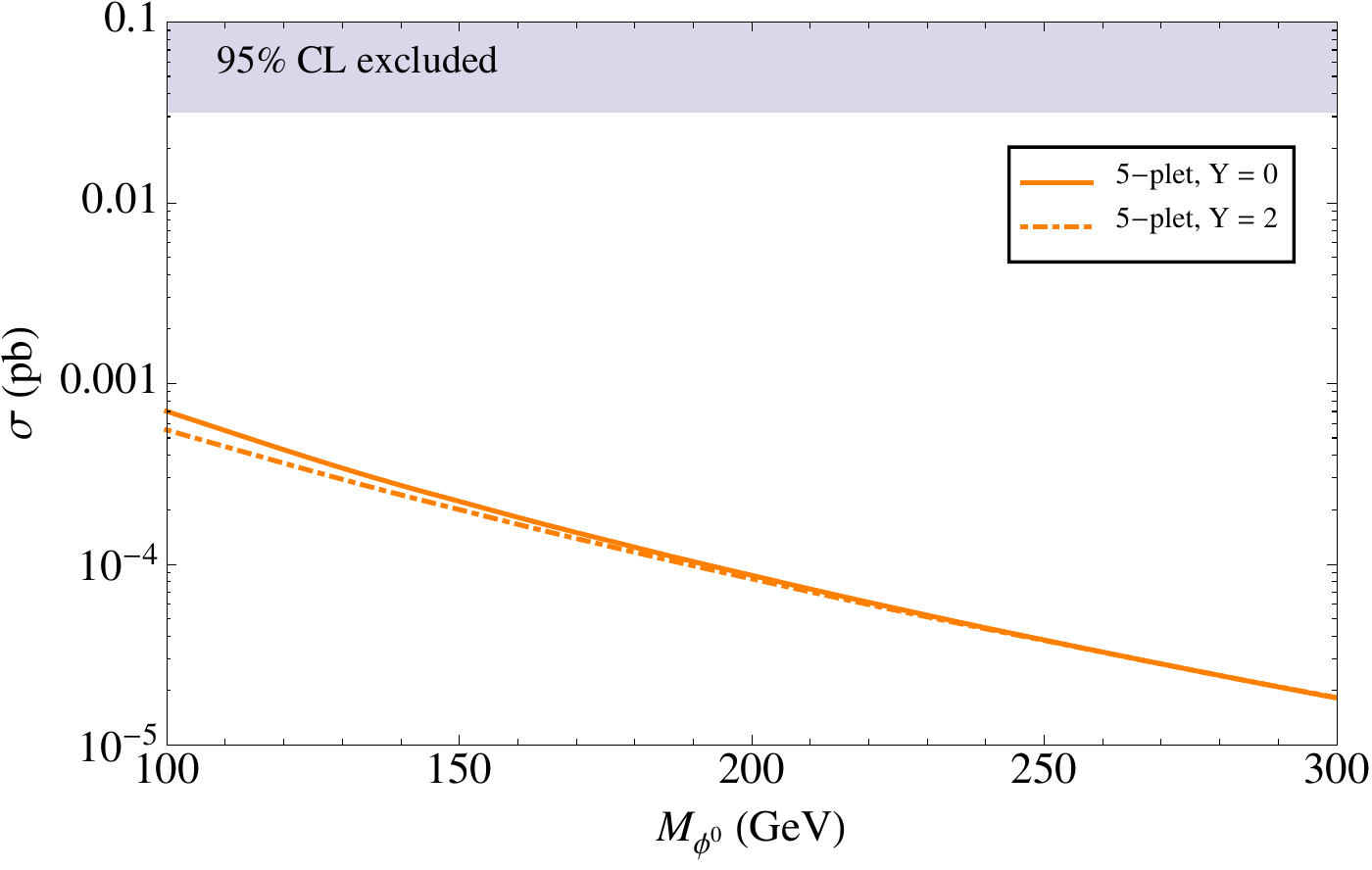}
  \end{center}
  \caption{The production cross section for a multiplet pair together with a
    monojet for our benchmark models. The cross sections (colored curves) are
    compared with a bound from the CMS monojet search~\cite{Chatrchyan:2012me}
    (gray area), which requires the reconstructed MET to be above 350~GeV.
    Following \cite{Chatrchyan:2012me}, we use a cut efficiency of 10\% relative to a Monte Carlo sample with MET$ > 200$~GeV. In order to get the detector level cross section with MET$ > 200$~GeV, we apply a 75\% efficiency to a Monte Carlo sample with parton level cut MET$ > 200$~GeV as suggested by Delphes simulation \cite{deFavereau:2013fsa}.}
  \label{fig:monojet}
\end{figure}

\begin{figure}
  \begin{center}
    \includegraphics[width=0.8\textwidth]{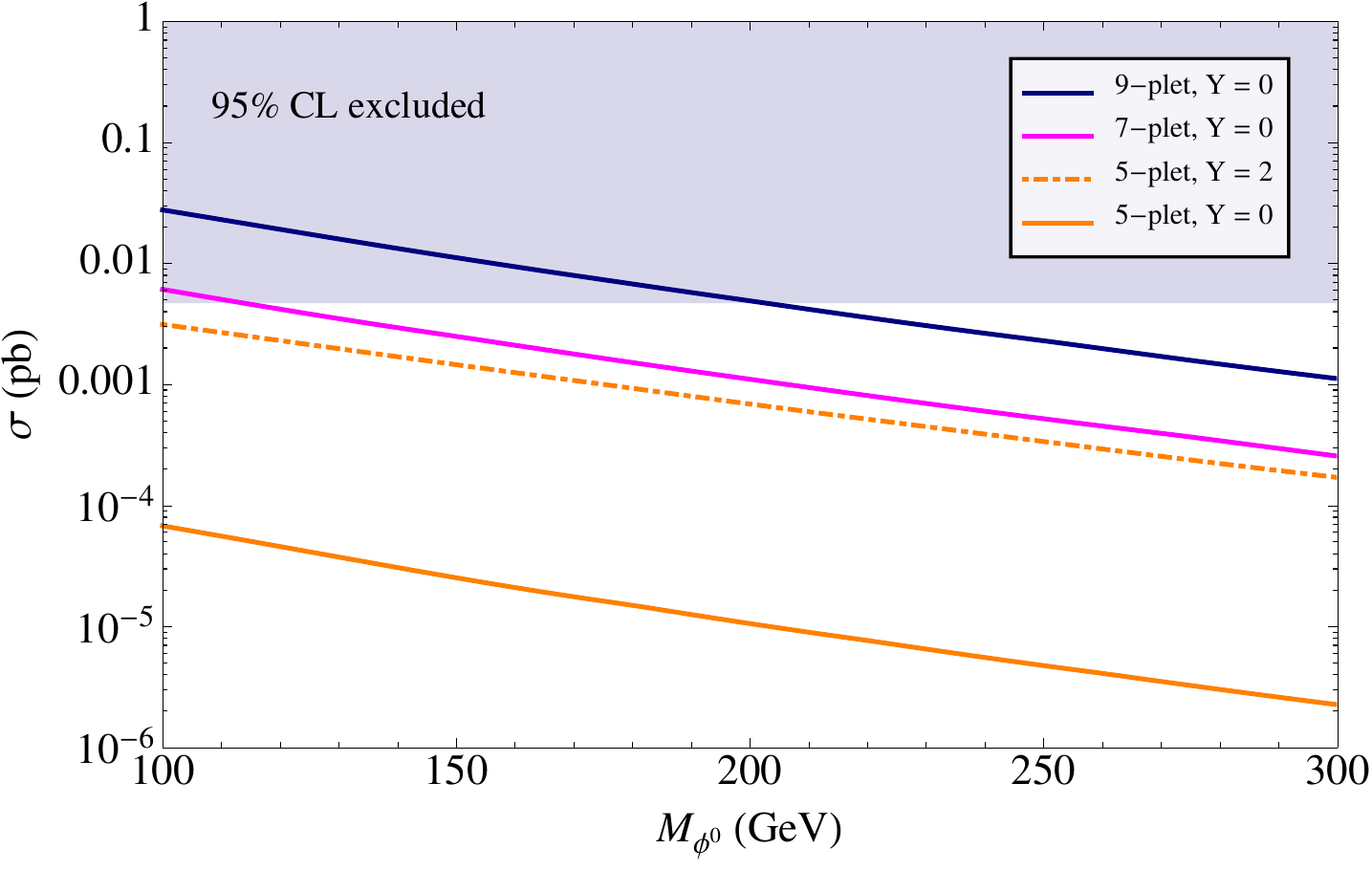}
  \end{center}
  \caption{The production cross section for a multiplet pair together with one photon for
    various multiplet quantum numbers. Based on the CMS 7~TeV monophoton
    search~\cite{Chatrchyan:2012tea}, we require $p_T > 125$ GeV and $|\eta| <
    1.5$ for the photon, and we take the signal efficiency to be 30\% relative
    to a parton level sample with photon $p_T > 125$~GeV and photon $\eta < 1.5$. The
    shaded region is excluded by the CMS search.}
  \label{fig:monophoton}
\end{figure}

Searches for a single jet or photon, accompanied by a significant amount of missing energy,
have recently received a lot of attention because they are able to constrain
the existence of new ``invisible'' particles in a relatively model-independent
way~\cite{Birkedal:2004xn, Beltran:2010ww, Goodman:2010yf, Bai:2010hh,
Goodman:2010ku, Fox:2011fx, Rajaraman:2011wf, Mambrini:2011pw, Fox:2011pm, Bartels:2012ui,
Fox:2012ee, Bartels:2012ex, Dreiner:2012xm, Chae:2012bq,
Aad:2012fw, ATLAS-CONF-2012-147, Chatrchyan:2012tea,
Chatrchyan:2012me}. In the models discussed here, for instance, the components of the
electroweak multiplet mediator $\phi$ are very difficult to observe directly at the
LHC, but their production is constrained by jet+MET and photon+MET searches.

For the monojet signature, the relevant diagrams are pair productions of the
multiplet together with a quark or gluon from initial state radiation. For our
benchmark points  the expected cross section is still about one order
of magnitude below the current LHC bound, as shown in \cref{fig:monojet}.

The production cross section for a $\phi$ pair together with a single photon is
significantly enhanced compared to the monojet case because hard photons can be
radiated not only from the initial state quarks, but also from the
$\phi^{n\pm}$ in the final state, which couple strongly to photons.  The
monophoton cross sections for our benchmark points, as well as two additional
cases with even larger $SU(2)$ representations, is shown in
\cref{fig:monophoton} and compared to the current limit from
CMS~\cite{Chatrchyan:2012tea}. While our benchmark models are still allowed with the current constraint, we expect them to be excluded by the monophoton searches in the near future. In addition, models with larger multiplet representations are starting to be in tension with the current monophoton constraint. 

Finally, we have also considered the signature of $\phi \phi^\dag + h$ production
through the operator $\phi^\dagger \phi H^\dag H$. The signature of this process---a Higgs boson
plus a lot of missing energy---is identical to the one for associated $Z+h$ production,
in which the $Z$ decays invisibly. For our $N=5$, $Y_\phi = 0$ benchmark point,
the cross section for $\phi \phi^\dag + h$ production at $\sqrt{s} = 8$~TeV
is 0.3~fb, while for the $N=5$, $Y_\phi = 2$ benchmark point, it is $8\times 10^{-2}$~fb.
Since the SM cross section for $Z+hh$ production
is $\sim 400$~fb, we do not expect to see any
modification of the Higgs plus missing energy event rate in the foreseeable future.

\section{Modification of Higgs boson decays}
\label{sec:higgs}

\begin{figure}
\begin{center}
\includegraphics[width=0.3\textwidth]{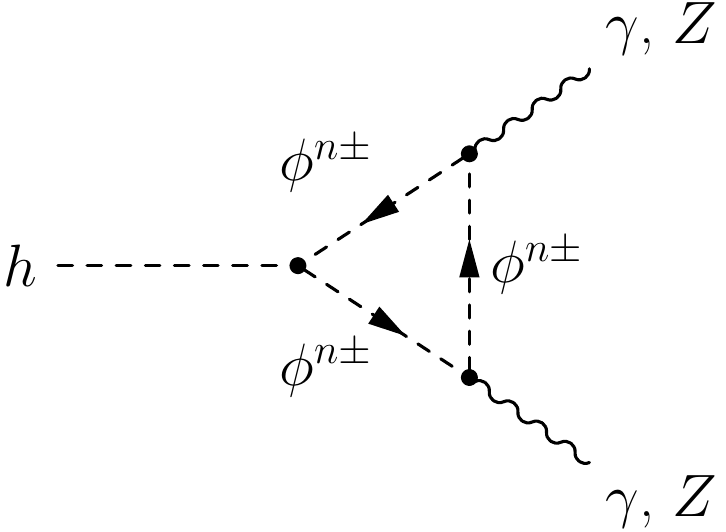}
\end{center}
\caption{Feynman diagram through which the new electroweak multiplet $\phi$ contributes to Higgs annihilation into $\gamma\gamma$ and $\gamma Z$ final states.}
\label{fig:h-annihilation-diagram}
\end{figure}

In a model with multiple scalar fields, the presence of ``Higgs portal"-type
operators is quite natural; indeed, as these are dimension-four operators consistent with
all of the other symmetries, they are difficult to forbid without ad-hoc assumptions.  The presence of the operator
 $\phi^\dagger \phi H^\dagger H$ can significantly modify 
decays of the Higgs boson to gauge bosons, especially $h\to \gamma\gamma$ and $h\to Z\gamma$ which arise in the SM only at loop level.

We first consider the $h \rightarrow \gamma \gamma$ decay width
\begin{align}
  \Gamma(h \rightarrow \gamma \gamma) = \frac{v^2}{16\pi M_H} |F^{\gamma\gamma}|^2 \,.
  \label{eq:h-gammagamma}
\end{align}
where  $v = 246$~GeV is the vacuum expectation value of the SM Higgs and $F^{\gamma\gamma}$ a dimensionless amplitude. 
At one loop this amplitude receives SM contributions from the $W$ boson loop, $F^{\gamma\gamma}_W$,  and from fermion loops, $F^{\gamma\gamma}_f$~\cite{Shifman:1979eb, Spira:1995rr, Djouadi:2005gi,Djouadi:2005gj}. In our model there is an additional contribution, $F^{\gamma\gamma}_\phi$, from the scalar multiplet $\phi$ running in the loop (see \cref{fig:h-annihilation-diagram}), so that
\begin{align}
  F^{\gamma\gamma} = F^{\gamma\gamma}_W + F^{\gamma\gamma}_f + F^{\gamma\gamma}_\phi \,.
\end{align}
Using analytic expressions from \cite{Djouadi:2005gi, Djouadi:2005gj} we have
\begin{align}
  |F^{\gamma\gamma}_W| &= 1.25 \times 10^{-3} \,,\qquad   |F^{\gamma\gamma}_f| = 2.75 \times 10^{-4} \,,
\end{align}
for $M_H = 125$~GeV, $M_W = 80.4$~GeV, and the top quark mass in
the $\MSbar$ renormalization scheme $M_t^{\MSbar} = 160$~GeV \cite{Beringer:1900zz}.
We neglect the contributions from fermions other
than the top quark, which is sufficient to reproduce the full SM result for the
partial width $\Gamma(h \rightarrow \gamma \gamma)$~\cite{Dittmaier:2011ti} to
within 2\%.  The new contribution $F^{\gamma\gamma}_\phi$ is given by~\cite{Shifman:1979eb,
Djouadi:2005gi, Djouadi:2005gj}
\begin{align}
  F^{\gamma\gamma}_\phi = \sum_s \frac{\alpha Q_s^2}{2\pi}
    \Big(\lambda_{\phi H} - \frac{1}{2} \lambda'_{\phi H} (T^3)_s \Big)
    \Big[1 - \beta_s f(\beta_s) \Big] \,,
  \label{eq:Fphi-gammagamma}
\end{align}
where the sum runs over the components $\phi_s$ of the electroweak multiplet $\phi$, and
the loop function $f(\beta_s)$ is given by \cref{eq:loop-function-f},
and $\beta_s \equiv 4M_{\phi_s}^2 / M_H^2$.  For our
benchmark models
$\beta_s > 1$.
Note that for $Y=0$, the term proportional to $\lambda'_{\phi H}$ vanishes.

\begin{figure}
\begin{center}
\includegraphics[width=0.7\textwidth]{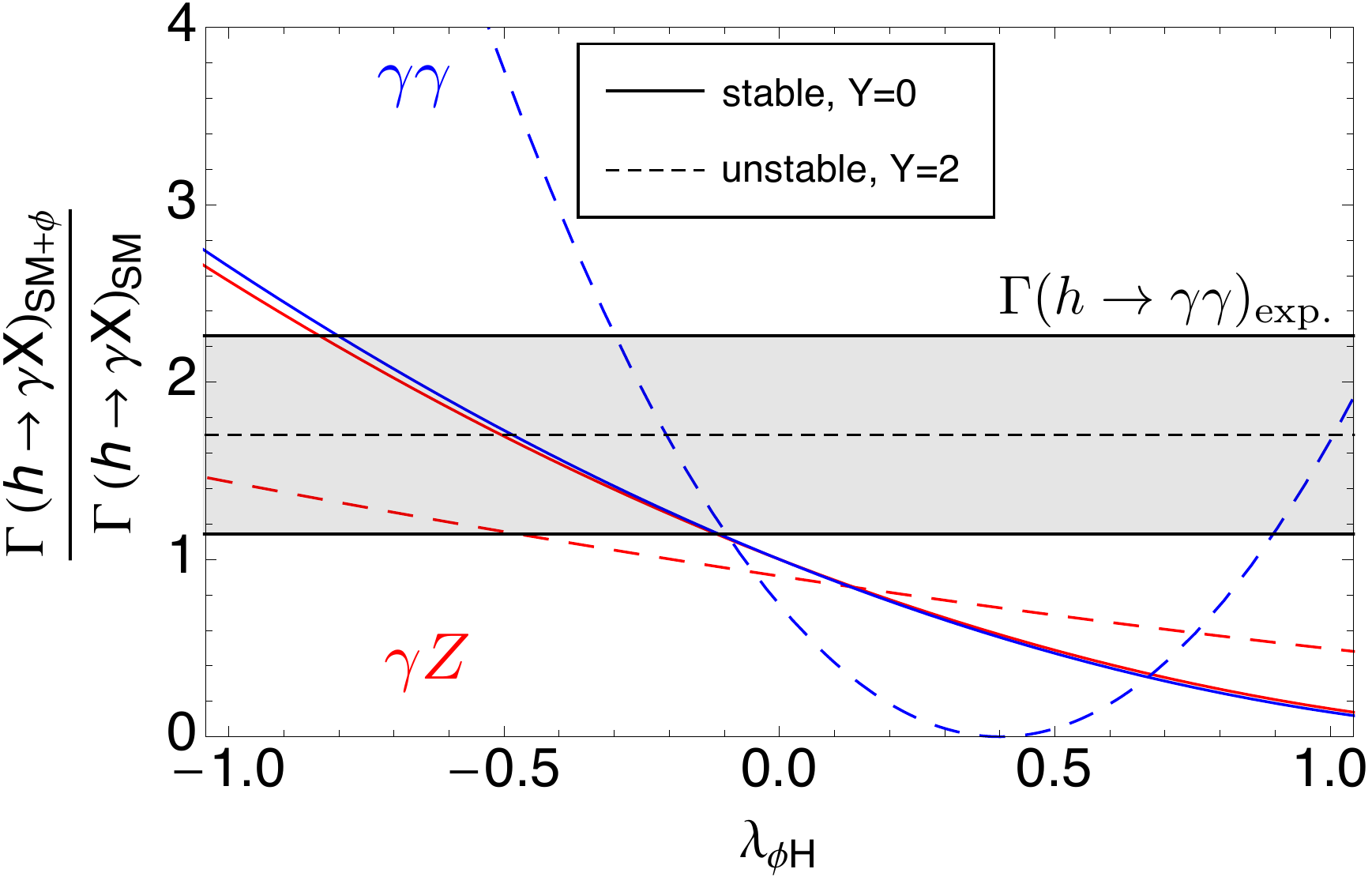}
\end{center}
\caption{Ratio of partial decay width of the Higgs boson in our model to the standard model width for decay modes $\gamma \gamma$ (blue) and $\gamma Z$ (red), as a function of the coupling $\lambda_{\phi H}$.  The solid and dashed curves correspond to the stable and unstable benchmark points of \cref{tab:benchmarkpoint}, respectively.  Experimental results for best-fit signal strength in the channel $h \rightarrow \gamma \gamma$ (dashed horizontal line) are taken from \cite{ATLASHiggsgammaNew,:2012gu} with errors added in quadrature (gray band).}
\label{fig:hAA}
\end{figure}

In the SM 
$F^{\gamma\gamma}_W$ and $F^{\gamma\gamma}_f$ interfere
destructively.
Furthermore, 
for positive $\lambda_{\phi H}$ and $\lambda'_{\phi H}$ we have
\begin{align}
  |F^{\gamma\gamma}|^2 = \big( |F^{\gamma\gamma}_V| - |F^{\gamma\gamma}_F|
                             - |F^{\gamma\gamma}_\phi|\big)^2.
\end{align}
The ratio of the resulting partial width $\Gamma(h \rightarrow \gamma \gamma)$
to the 
SM value as a function of $\lambda_{\phi H}$ is shown in
\cref{fig:hAA}. We see that the decay $h \to \gamma\gamma$ can be substantially
enhanced or suppressed, depending on the sign of the coupling constants $\lambda_{\phi H}$ and $\lambda_{\phi H}'$. Note that the  other phenomenology discussed so far is to a large extent decoupled from the value of $\lambda_{\phi H}$, so that ${\mathcal O}(1)$ effects in $h \to \gamma\gamma$ are possible without affecting anything else.  In particular, there is no clear prediction for the size of the deviation in $h \to \gamma\gamma$ based on observation of the Fermi gamma line, beyond the generic expectation that ${\mathcal O}(1)$ deviation is expected for natural values of $\lambda_{\phi H}$.

In a similar way, the related loop-induced decay $h \to \gamma
Z$ is affected by the 
new multiplet $\phi$. 
The decay rate is given by
\begin{align}
  \Gamma(h \rightarrow \gamma Z) = \frac{v^2}{16\pi M_H}
    \bigg(1 - \frac{M_Z^2}{M_H^2}\bigg)^3 |F^{\gamma Z}|^2 \,,
  \label{eq:h-gammaZ}
\end{align}
where the amplitude $F^{\gamma Z}$ receives contributions from $W$
loops, fermion loops, and $\phi$ loops,
\begin{align}
  F^{\gamma Z} = F^{\gamma Z}_W + F^{\gamma Z}_f + F^{\gamma Z}_\phi \,.
\end{align}
Analytic expressions for these 
can be found,
e.g.,
in refs.~\cite{Djouadi:2005gi, Djouadi:2005gj}. 
In the SM we have
\begin{align}
  |F^{\gamma Z}_W| &= 2.63 \times 10^{-3} \,, \qquad
  |F^{\gamma Z}_f| = 1.41 \times 10^{-4} \,.
\end{align}
The new physics contribution is
\begin{align}
  F^{\gamma Z}_\phi(\beta_s, \gamma_s) &= \sum_s 
    \frac{\sqrt{2 \alpha \alpha_2} Q_s ((T^3)_s - s_W^2 Q_s)}{\pi}
    \Big(\lambda_{\phi H} - \frac{1}{2} \lambda'_{\phi H} (T^3)_s \Big) \nonumber\\
  &\quad \times
      \bigg[
          \frac{\gamma_s}{2(\beta_s-\gamma_s)}
        + \frac{\beta_s \gamma_s^2}{2(\beta_s-\gamma_s)^2} \big[ f(\beta_s) - f(\gamma_s) \big]
        + \frac{\beta_s \gamma_s}{(\beta_s-\gamma_s)^2}    \big[ g(\beta_s) - g(\gamma_s) \big]
      \bigg]\,.
  \label{eq:Fphi-gammaZ}
\end{align}
Here, as in \cref{eq:Fphi-gammagamma}, the sum runs over the components of
$\phi$, while 
$\beta_s \equiv 4M_{\phi_s}^2 / M_H^2$, $\gamma_s \equiv 4M_{\phi_s}^2 / M_Z^2$,
and the loop functions $f(\beta_s)$ and $g(\gamma_s)$ have been defined in
\cref{eq:loop-function-f} and \cref{eq:loop-function-g}. Numerically, the new physics contribution is of the same order as the SM contribution for $\lambda_{\phi H}\sim {\mathcal O}(1)$; see \cref{fig:hAA}.

Similar loop contributions correct the $h \to WW$ and
$h \to ZZ$ branching ratios.  However, since these processes receive tree level SM contributions,
the relative corrections from the new multiplet are small.  
They interfere destructively with the tree-level amplitude, leading to
a slight reduction in the partial decay widths for $h\to WW$ and $h\to ZZ$.  Numerical evaluation of the loop
diagrams using {\tt FeynArts} for our benchmark points yields corrections on the order
of a few percent.

Modifications of the other Higgs decay modes are negligible, since
$\phi$ has no direct coupling to any of the SM fermions.  Invisible
decays of the Higgs into $\phi$ or $\chi$ are forbidden kinematically for the
regions of parameter space we consider.

\section{Conclusion}
\label{sec:conc}

In conclusion, we have investigated the possible implications of gamma ray
lines in astrophysical dark matter searches for the LHC.  Motivated by the
tenative hints for a line-like signal at $\sim$~130~GeV in Fermi-LAT data, we
have focused on a class of secluded DM models, in which the DM particle $\chi$ couples to Standard Model fermions and gauge bosons
through loops of an intermediate particle $\phi$.  We have considered in detail
the case where both $\chi$ and $\phi$ are vev-less scalars, but we expect our
conclusions to be valid also in a more general context. We have moreover
assumed that $\phi$ belongs to a large representation ($\mathbf{N} > 3$) of the
weak $SU(2)$ gauge group.

Among the models proposed to explain the Fermi-LAT signal (if it stands up to
further experimental scrutiny), this scenario has several advantages:
1) For natural, untuned values of the coupling constants, it can
explain relatively large ($\ev{\sigma v} \sim 10^{-27}$--$10^{-26}$~cm$^3$/sec)
DM annihilation cross sections to $\gamma \gamma$ and/or $\gamma Z$
final states.  Both of these final states lead to monoenergetic features in the
astrophysical gamma ray spectrum. The key is that some of the component
fields of $\phi$ carry several units of electric charge, which significantly
enhances the DM coupling to photons. 2) All coupling constants are
perturbative at experimentally relevant energy scales. 3) The model is well
compatible with the observed DM relic abundance in the Universe. 4)
DM annihilation to $WW$, $ZZ$, and fermion-antifermion final states is
small. This is important because these final states are tightly constrained by
searches for the broad excess they would induce in the Fermi-LAT data. 5) DM--nucleon scattering cross sections are compatible with current
constraints from direct DM searches, but are testable in future
experiments.

Turning to the LHC phenomenology, we found that the scenarios we consider have
very characteristic signatures, but are still largely unconstrained at present.
While the members of the mediator multiplet
$\phi$ can be copiously produced due to their large electroweak couplings, they
are difficult to observe because their cascade decays down to the lightest
component field are typically very soft. The reason is that, unless there are
large, isospin-dependent couplings to the Higgs, the mass splittings among them
arise only from higher-order electroweak corrections. Thus, the energies of the
decay products can easily be well below the trigger thresholds of ATLAS and
CMS. The lightest component field, which we take to be the neutral one,
$\phi^0$, in turn, is invisible due to its vanishing electric charge. In view
of this, $\phi$ production can contribute to final states with
large missing transverse energy, for instance monophoton + MET. Here, the
probability for radiating an extra hard photon in $pp \to \phi\bar\phi$ is
enhanced by the large electric charge of the component fields
of $\phi$. We have found that monophoton + MET searches are already beginning to
constrain some regions of parameter space, and will be able to probe much
larger regions, including one of our two benchmark points, in the future.

A second place in which the multiplet $\phi$ can leave its footprint at the LHC
is the Higgs sector. Higgs boson decays to $\gamma\gamma$ and
$\gamma Z$ receive extra contributions from diagrams involving $\phi$ loops,
and these extra contributions can either enhance or suppress the corresponding
Higgs branching ratios. The LHC data on $h \to \gamma\gamma$ thus already
provides loose constraints on the $\phi$--$h$ couplings, as shown in
\cref{fig:hAA}.

Finally, if the mass splittings among the components of $\phi$ are very small,
some of them can be sufficiently long-lived to yield anomalous charged tracks
that can be detected in future searches using the ATLAS and CMS inner detectors.
Other signatures that might provide promising starting points for future work
include photon + MET + X. With optimized cuts, these signatures could efficiently
exploit the large $\phi$--$\gamma$ couplings to improve the signal-to-background
ratio.

\section*{Acknowledgments}

We thank Jared Evans, Marco Nardecchia, and Felix Yu for useful discussions
during the preparation of this manuscript. We are especially grateful to
the authors of~\cite{Nelson:2013pqa}, in particular Randel Cotta,
for pointing out a mistake in a previous version of figs.~\ref{fig:monojet}
and \ref{fig:monophoton}.
Fermilab is operated by Fermi
Research Alliance, LLC, under Contract DE-AC02-07CH11359 with the United States
Department of Energy. JZ was supported in part by the U.S.\ National Science
Foundation under CAREER Grant PHY-1151392.  RP is supported by the NSF under
grant PHY-0910467. JK is grateful to the Galileo Galilei Institute, Firenze,
Italy for warm hospitality and support during part of this work.

\appendix

\section{SU(2) interactions}
\label{sec:generators}

In this appendix we give the normalization of the $SU(2)_L$ generators used in the paper and then also write out explicitly the gauge interactions. For the generators $T_N^a$ of the representation $\mathbf{N}$ of the algebra $su(2)$, we fix the normalization by insisting that for any representation, the eigenvalues of $T^3$ (and thus the electric charge of the multiplet scalars) differ by integers.  Thus, we have explicitly in the basis with $T^3_N$ diagonal
\beq
(T_N^3)_{mn} = \left(\frac{N+1}{2}-m\right) \delta_{mn},
\eeq
where $m \in [1,N]$.  Since $su(2) \cong so(3)$, the other generators of representation $\mathbf{N}$ can be obtained from the familiar angular-momentum ladder operators for spin $j = (N-1)/2$,
\begin{align}
(T_N^+)_{mn} &= \sqrt{m(N-m)} \delta_{m+1,n}, \qquad (T_N^-)_{mn} = \sqrt{(m-1)(N-m+1)} \delta_{m-1,n},
\end{align}
and the relation $T_N^{\pm} \equiv T_N^1 \pm iT_N^2$.  It is easily verified that the three generators $T_N^a$ satisfy the defining relation
\beq
[T_N^a, T_N^b] = i \epsilon^{abc} T_N^c.
\eeq
In the fundamental representation $\mathbf{2}$, these generators match on to the usual Pauli matrices, $T^a_2 = \sigma^a / 2 \equiv \tau^a$.

For the calculation of the $T$-parameter in \cref{sec:constrainEW}, we make use of the relations
\begin{align}
(T_N^1)^2_{ss} &= (T_N^1)_{ss'} (T_N^1)_{s's} \nonumber \\
&= \frac{1}{4} \left[ s(N-s) + (s-1)(N-s+1) \right].
\end{align}
and
\begin{align} \label{eq:T1square}
(T^1_{N,ss'})^2 &= \left(\frac{1}{2} (\sqrt{s(N-s)} \delta_{s+1,s'} + \sqrt{(s-1)(N-s+1)} \delta_{s-1,s'})\right)^2 \nonumber \\
&= \frac{1}{4} s(N-s) \delta_{s+1,s'} + \frac{1}{4}(s-1)(N-s+1) \delta_{s-1,s'}.
\end{align}

Using this normalization the  Lagrangian \cref{Lagr-gauge} is, in the 
gauge-field mass eigenstate basis,
\begin{align}
\mathcal{L} &\supset i[\phi_i^\dagger \partial^\mu \phi_j - (\partial^\mu \phi_i^\dagger) \phi_j]\left[\frac{g}{\sqrt{2}} (W_\mu^+ (T_N^+)_{ij} + W_\mu^- (T_N^-)_{ij}) + \delta^{ij} \left( \frac{g}{\cos \theta_w} Z_\mu (T_N^3 - \sin^2 \theta_w Q)_{ij} + eA_\mu Q\right) \right] \nonumber \\
&+ \phi_i^\dagger \phi_j \left[ \frac{g^2}{4} W_\mu^\pm W^{\mu,\mp} \left\{T_N^+, T_N^-\right\}_{ij} + \delta^{ij} \left( Z_\mu Z^\mu \frac{g^2}{\cos^2 \theta_w} (T_N^3 - \sin^2 \theta_w Q)^2  + A_\mu A^\mu Q^2 e^2 \right. \right. \nonumber \\
&\left. \left. \hspace{10mm} + 2A_\mu Z^\mu eQ \frac{g}{\cos \theta_w} (T_N^3 - \sin^2 \theta_w Q) \right)  + W_\mu^\pm A^\mu \frac{eg}{\sqrt{2}} \left(\{T_N^\pm, T_N^3\}_{ij} + 2(T_N^\pm)_{ij} Y_\phi\right)  \right. \nonumber \\
&\left. \hspace{10mm} + W_\mu^\pm Z^\mu \frac{g^2}{\sqrt{2} \cos \theta_w} \left( \{T_N^\pm, T_N^3\}_{ij} \cos^2 \theta_w - 2(T_N^\pm)_{ij} Y_\phi \sin^2 \theta_w\right)  \right]. \label{eq:gaugeL}
\end{align}

The $\lambda'_{\phi H}$ interaction term in \cref{Lagr-phi} has an unusual form, and at first glance may not appear to be gauge invariant.  We can demonstrate its invariance by performing an arbitrary SU$(2)$ gauge transformation:
\beq
(\phi^\dagger T_N^a \phi) (H^\dagger \tau^a H) \rightarrow \sum_{b,d} \left(\phi^\dagger e^{i \theta^b (T_N^b)^\dagger} T_N^a e^{-i \theta^b T_N^b} \phi\right) \left( H^\dagger e^{i \theta^d (\tau^d)^\dagger} \tau^a e^{-i \theta^d \tau^d} H\right).
\eeq
Expanding to first order in the parameter $\theta^a$, the bilinears transform as
\beq
\phi^\dagger T_N^a \phi \rightarrow \sum_b \phi^\dagger (1 + i\theta^b T_N^b) T^a (1 - i\theta^b T_N^b) \phi = (\delta^{ac} - \theta^b f^{abc}) \phi^\dagger T_N^c \phi, 
\eeq
and similarly for $H^\dagger \tau^a H$.  The bilinear itself is not gauge invariant, but for the combination we find
\beq
(\phi^\dagger T_N^a \phi) (H^\dagger \tau^a H) \rightarrow (\phi^\dagger T_N^a \phi) (H^\dagger \tau^c H) \left[ \delta^{ac} - \theta^b f^{cba} - \theta^b f^{abc} \right],
\eeq
and the extra terms vanish by the total antisymmetry of the structure constants $f^{abc}$.

\section{Dark matter direct detection}
\label{sec:app:direct detection}

In this appendix we collect the analytical results for DM scattering on nuclei for our models where DM interacts with the visible sector through electroweak multiplets and the Higgs. The heavy fields, $\phi$, $W$, $Z$ and Higgs are integrated out and one matches onto the Effective Field Theory (EFT) operator basis \eqref{ops-basis-EFT}.\footnote{Note that at this stage also the top quark should be integrated out. In what follows we will treat the top contributions to DM--nucleon scattering only very approximately.} These EFT operators induce spin independent DM--nucleon scattering, \cref{xsec-DM-nucleon}.

The only tree level contribution is due to a single Higgs exchange, giving
\beq
C_q^\text{tree}= -\lambda_{\chi H}\frac{M_\phi^2}{M_H^2}.
\eeq
This contribution is absent in our benchmark points, where we set $\lambda_{\chi H}=0$, but  could in principle saturate the present DM-nucleon direct detection bounds if  $\lambda_{\chi H}\sim 3\cdot 10^{-2}$. For such small values of $\lambda_{\chi H}$ the effects on annihilation cross section and early cosmology are atill very small, though.

The $C_F$ Wilson coefficients is first nonzero at 1~loop, where $\phi$ fields run in the loop, giving
\beq
C_F=\frac{\alpha}{24 \pi} \lambda_{\chi \phi} \sum_\phi Q_\phi^2,
\eeq
with $Q_\phi$ the charges of the $\phi$ field components, and the sum runs over all the components in the multiplet (for simplicity we have treated the masses of the $\phi$ components as degenerate). For the 5-plet benchmark model with $Y_\phi = 0$, we have thus $C_F = 5 \alpha \lambda_{\chi \phi}/(12\pi)=9 \cdot 10^{-4}$, and for the 5-plet with $Y_\phi = 2$ we find $C_F=5 \alpha \lambda_{\chi \phi}/(4\pi)=1 \cdot 10^{-3}$.

The $C_q$ Wilson coefficient also receives the 2-loop contributions with $\phi$ and $W$, $Z$, $\gamma$ in the loops. We calculate these contributions in the approximation where $M_\phi \gg m_{W,Z}$. In that case we can first integrate out the $\phi$ fields and match onto an EFT with the operator $\chi^2 F_{\mu\nu}F^{\mu\nu}$ from \eqref{ops-basis-EFT}, as well as the operators 
\beq
{\cal L}_{eff}\supset \frac{{\cal C}_{Z}}{M_\phi^2}\chi^2 Z_{\mu\nu}Z^{\mu\nu}+\frac{{\cal C}_{Z\gamma}}{M_\phi^2}\chi^2 Z_{\mu\nu}F^{\mu\nu}+\frac{{\cal C}_{W}}{M_\phi^2}\chi^2 W_{\mu\nu}^+W^{-\mu\nu} \,.
\eeq
Here, we have defined $Z_{\mu\nu}\equiv \partial_\mu Z_\nu-\partial_\nu Z_\mu$ and $W^\pm_{\mu\nu}\equiv \partial_\mu W^\pm_\nu -\partial_\nu W^\pm_\mu$. Note that electroweak symmetry is already broken in this EFT. The Wilson coefficients are 
\begin{align}
{\cal C}_Z&=\frac{\alpha}{24\pi}\frac{\lambda_{\chi \phi}}{(s_W c_W)^2}\sum_\phi (T^3_\phi-s_W^2 Q_\phi)^2 \,, \\
{\cal C}_W&=\frac{\alpha}{24\pi}\frac{\lambda_{\chi \phi}}{s_W^2}{\text{Tr}}(T^+ T^-) \,,\\
{\cal C}_{Z\gamma}&=\frac{\alpha}{24\pi}\frac{\lambda_{\chi \phi}}{s_Wc_W}\sum_\phi (T^3_\phi-s_W^2 Q_\phi) Q_\phi \,,
\end{align}
with the sums again running over the components of $\phi$.
For the $N=5$, $Y_\phi = 0$ benchmark model ${\cal C}_Z = 3\times 10^{-3}$, ${\cal C}_W = 8\times 10^{-3}$, ${\cal C}_{Z\gamma} = 2\times 10^{-3}$. For the $N=5$, $Y_\phi = 2$ benchmark model, the Wilson coefficients are ${\cal C}_Z = 2\times 10^{-3}$, ${\cal C}_W = 4\times 10^{-3}$, ${\cal C}_{Z\gamma} = 3\times 10^{-4}$.

Integrating out $W$ and $Z$, the one loop contributions with $ZZ$, $WW$ and $Z\gamma$ running in the loop give 
\beq
C_q=C_q^Z+C_q^W+C_q^{Z\gamma},
\eeq
with 
\begin{align}
C_q^Z&=\frac{\alpha}{4\pi}\frac{{\cal C}_Z}{(s_W c_W)^2}\big[-2 Q_q^2 s_W^4 + 2 Q_q s_W^2 T_q^3 + 3 (T_q^3)^2 + 12 Q_q^2 s_W^2 \log(M_Z^2/\mu^2)\big[T_q^3 - Q_q s_W^2\big]\big],\\
C_q^W&=\frac{3 \alpha}{8 \pi}\frac{{\cal C}_W}{s_W^2},\\
C_q^{Z\gamma}&=\frac{\alpha}{4\pi}\frac{{\cal C}_{Z\gamma} Q_q}{s_W c_W}\big[-5+6\log(M_Z^2/\mu^2)\big]\big[2 Q_q s_W^2 - T_q^3\big].
\end{align}

\section{Calculation of the $S$- and $T$-parameters}
\label{sec:Sparam}

For the $S$-parameter contribution from the scalar multiplet $\phi$, we need to compute the vacuum polarization amplitude $\Pi'_{3Y}(0)$.  At one loop, only a single type of Feynman diagram contributes, with an intermediate loop of scalar particles (if $\phi$ had a vev, there would be additional contributions with internal gauge-boson propagators.)  Labeling the external momentum as $p$ and the loop momentum as $k$, the amplitude is given by the expression
\beq
i \Pi_{3Y}^{\mu \nu} = \sum_{s,s'} \int\! \frac{d^4 k}{(2\pi)^4} \frac{i}{k^2 - M_s^2 + i\epsilon} (-igT^3_{N,ss'} (-2k-p)^\mu) \frac{i}{(p+k)^2 - M_{s'}^2 + i\epsilon} (-ig'Y (2k+p)^\nu).
\eeq
Using the standard Feynman parameterization, we can shift the integration momentum and rewrite:
\beq
\Pi_{3Y}^{\mu \nu} = i gg' Y \sum_{s} T^3_{N,ss} \int_0^1 dx \int\! \frac{d^4 \ell}{(2\pi)^4} \frac{4 \ell^\mu \ell^\nu + (1-2x)^2 p^\mu p^\nu}{[\ell^2 - (M_s^2 - x(1-x)p^2)]^2}.
\eeq
The second term proportional to $p^\mu p^\nu$ does not contribute to the transverse vacuum polarization (and thus to $S$), so we drop it.  As for the first term, Lorentz invariance allows us to replace $\ell^\mu \ell^\nu \rightarrow \frac{1}{4} \ell^2 g^{\mu \nu}$.  
Evaluating the momentum integral in dimensional regularization, we have
\beq
\Pi_{3Y}(p^2) = \frac{-gg'Y}{8\pi^2} \sum_s T^3_{N,ss} \int_0^1 dx\ \Delta_s (1+E - \log[ \Delta_s / \mu^2]),
\eeq
where $\Delta_s \equiv M_s^2 - x(1-x)p^2$, $E \equiv 2/\epsilon - \gamma + \log(4\pi) - \log(\mu^2)$, and $\mu$ is the renormalization mass scale.  It is clear at this point that if there is no mass splitting within the multiplet, then the amplitude is proportional to $\tr(T^3_N) = 0$, and there is no contribution to the $S$-parameter.  To convert to $S$, we need to take the ``derivative" at $p^2 = 0$, i.e.
\beq
S = -16\pi \Pi'_{3Y}(0) = -16\pi \lim_{p^2 \rightarrow 0} \left[ \frac{1}{p^2} \big(\Pi_{3Y}(p^2) - \Pi_{3Y}(0)\big) \right].
\eeq
Making use of the identity
\beq
\lim_{p^2 \rightarrow 0} \frac{\log[(M_s^2 - x(1-x)p^2)/M_s^2]}{p^2} = \frac{-x(1-x)}{M_s^2},
\eeq
we find that
\begin{align}
S &= -\frac{gg'Y}{3\pi} \sum_s T^3_{N,ss} \left( 2+E + \log(M_s^2 / \mu^2) \right).
\end{align}
The first term vanishes due to the trace over $T^3_N$, so the only contribution to $S$ is due to the logarithm.  Since the states of $\phi$ come in pairs of equal but opposite $T^3_N$ eigenvalues $I_3$, the dependence on the scale $\mu$ cancels, and we are left with our final expression,
\beq
S = -\frac{4 \alpha Y}{3 \sin \theta_w \cos \theta_w} \sum_{I_3 > 0} \left[ I_3 \log \left(\frac{M_{\phi,Y_\phi+I_3}^2}{M_{\phi,Y_\phi-I_3}^2}\right) \right].
\eeq

The calculation of the $T$-parameter is somewhat more involved, since it depends on the correlation functions directly and not just their derivatives.  This means that an additional diagram contributes to $T$, arising from the four-boson interaction $\phi^\dagger \phi A_{\mu,a} A^{\mu,a}$.  Furthermore, the loop coming from three-boson vertices can now have two distinct species of $\phi$ in the loop, due to the off-diagonal structure of $T^1_N$.  We thus have
\begin{align}
i\Pi_{ab}^{3,\mu\nu}(p) &= \sum_{s,s'} \int\!\frac{d^4 k}{(2\pi)^4} \frac{i}{k^2 - M_s^2 + i\epsilon} (-igT^a_{N,ss'}) (-2k-p)^\mu \frac{i}{(p+k)^2 - M_{s'}^2 + i\epsilon} (-igT^b_{N,s's}) (2k+p)^\nu, \\
i\Pi_{ab}^{4,\mu\nu}(p) &= \sum_{s} \int\!\frac{d^4 k}{(2\pi)^4} \frac{i}{k^2 - M_s^2 + i\epsilon} ig^2 g^{\mu \nu} \{T^a_N, T^b_N\}_{ss}.
\end{align}
Once again carrying out the momentum integral in dimensional regularization, taking the transverse part, and evaluating at $p^2 = 0$, we find
\begin{align}
\Pi_{aa}^{3}(0) &= -\frac{g^2}{8\pi^2} \sum_{s,s'} (T^a_{N,ss'})^2 \int_0^1 dx [(xM_s^2 + (1-x) M_{s'}^2)(1+E-\log\left(\frac{xM_s^2 + (1-x)M_{s'}^2}{\mu^2}\right))], \\
\Pi_{aa}^{4}(0) &= \frac{g^2}{8\pi^2} \sum_s (T^a_N)^2_{ss} M_s^2 [1 + E - \log(M_s^2 / \mu^2)].
\end{align}
Here $(T^a_{N,ss'})^2$ is the square of the matrix element $ss'$ of generator $T^a_N$, not to be confused with the matrix element $ss'$ of the squared generator $(T^a_N)^2$.  Making use of our explicit representation of the group generators, it can be verified that the $1/\epsilon$ divergence and scale dependence completely cancel in the difference $\Pi_{11}(0) - \Pi_{33}(0)$.  The leftover contribution comes from $\Pi_{11}^{3}(0)$, and is equal to
\beq
\Pi_{11}(0) - \Pi_{33}(0) = -\frac{g^2}{8\pi^2} \sum_{s,s'} (T^1_{N,ss'})^2 \left[ \frac{1}{4} (M_s^2 + M_{s'}^2) - \frac{M_{s'}^4 \log(M_s^2 / M_{s'}^2)}{2(M_s^2 - M_{s'}^2)}\right].
\eeq
Making use of the identity \cref{eq:T1square}, and the definition $T = 4\pi / (s^2 m_W^2) [\Pi_{11}(0) - \Pi_{33}(0)]$, we have finally
\begin{align}
T &= \frac{-\alpha}{2 \sin^4 \theta_w M_W^2} \sum_{s,s'} \left[  \frac{1}{4} s(N-s) \delta_{s+1,s'} + \frac{1}{4}(s-1)(N-s+1) \delta_{s-1,s'} \right] \nonumber \\
&\times \left[ (M_s^2 + M_{s'}^2) - \frac{2M_s^4}{M_s^2 - M_{s'}^2} \log (M_s^2 / M_{s'}^2) \right].
\end{align}

\section{Tree level annihilation cross section}
\label{sec:treelevel}

If the coupling of DM to the Higgs is non-zero, it can annihilate at tree level annihilation to $W^+W^-$, $ZZ$ and fermion--antifermion final states which contributes to the astrophysical continuum photon flux. The cross sections for DM annihilation to gauge bosons are given by
\begin{eqnarray}
\sigma_{W^+W^-}&=&\frac{\lambda_{\chi H}^2}{\pi}\frac{M_W^4}{s(s-M_H^2)^2}\left(3-\frac{s}{M_W^2}+\frac{s^2}{4M_W^4}\right)\sqrt{\frac{s-4M_W^2}{s-4m_\chi^2}},\\
\sigma_{ZZ}&=&\frac{\lambda_{\chi H}^2}{2\pi}\frac{M_Z^4}{s(s-M_H^2)^2}\left(3-\frac{s}{M_Z^2}+\frac{s^2}{4M_Z^4}\right)\sqrt{\frac{s-4M_Z^2}{s-4m_\chi^2}},
\end{eqnarray}
where $M_W$ ($M_Z$) is the mass of $W$ ($Z$) boson. 

This gives $\ev{\sigma v_\text{rel}}_{W^+W^-}= \lambda_{\chi H}^2\times (2
\cdot10^{-23}\ \text{cm}^3/\text{s})$, $\ev{\sigma v_\text{rel}}_{ZZ} =
\lambda_{\chi H}^2\times (1 \cdot10^{-23}\ {\text{cm}}^3/\text{s})$ for our
stable benchmark point ($N = 5$, $Y_\phi = 0$) and $\ev{\sigma
v_\text{rel}}_{W^+W^-} = \lambda_{\chi H}^2\times (3 \cdot10^{-23}\
{\text{cm}}^3/\text{s})$, $\ev{\sigma v_\text{rel}}_{ZZ} = \lambda_{\chi
H}^2\times (1 \cdot10^{-23}\ {\text{cm}}^3/\text{s})$ for the unstable case
($N=5$, $Y_\phi=2$). Using the continuum photon flux bound from the galactic
center $\sim 3\cdot 10^{-26}\ \text{cm}^3/\text{s}$, this bounds $\lambda_{\chi
H}\lesssim 0.03$.

\bibliographystyle{apsrev}
\bibliography{./gamma-multiplet}

\end{document}